\documentclass[useAMS,usenatbib,usegraphicx]{mn2e}
\usepackage{psfig}
 
\def\apj{ApJ}%
\def\apjl{ApJ}%
\def\apjs{ApJS}%
\def\aap{A\&A}%
\def\mnras{MNRAS}%
\def\pasp{PASP}%

\def\ww{luminosity-weighted }
\begin{document}
 \topmargin=-1.3cm
\title[Nearby early-type galaxies with ionized gas.]{Nearby early-type galaxies with ionized gas. The UV emission from  GALEX observations\thanks{Based on GALEX  observations: GI3-0087 PI R. Rampazzo}}
\author[A. Marino et al.]{A. Marino$^{1}$\thanks{E-mail:amarino@pha.jhu.edu}, R. Rampazzo$^{2}$, L. Bianchi$^{1}$, F. Annibali$^{2}$, A.Bressan$^{2,5}$, L.M. Buson$^{2}$,\newauthor M.S. Clemens$^{2}$, P. Panuzzo$^{3}$, W.W. Zeilinger$^{4}$ \\  
$^{1}$  Dept. of Physics and Astronomy, Johns Hopkins University, 3400 North Charles
           Street, Baltimore, MD 21218  USA\\
$^{2}$ INAF Osservatorio Astronomico di Padova, vicolo dell'Osservatorio~5, I-35122  Padova,
Italy\\
$^{3}$ DSM/Irfu/Service d'Astrophysique, CEA Saclay, 91191 Gif sur Yvette Cedex, France\\
$^{4}$ Institut f\" ur Astronomie der Universit\" at  Wien, T\" urkenschanzstra$\ss$e 17, A-1180 Wien, Austria\\  
$^{5}$ SISSA, Via Beirut 4, I-34014 Trieste - Italy\\
}
\date{Accepted. Received}

\pagerange{\pageref{firstpage}--\pageref{lastpage}} \pubyear{2010}

\maketitle

\label{firstpage}

\begin{abstract}
 
 We present GALEX far-ultraviolet (FUV, $\lambda_{eff}$=1538 \AA)  and near-ultraviolet (NUV, $\lambda_{eff}$=2316 \AA) surface photometry 
 of 40  early-type galaxies (ETGs) selected from a wider sample of 65 nearby ETGs showing emission 
 lines in  their optical spectra.  
 We derive FUV  and NUV  surface brightness profiles, (FUV-NUV)
 colour profiles  and D$_{25}$  integrated magnitudes. 
 We extend the photometric study to the optical {\it r}  
 band from SDSS imaging for 14 of these ETGs.  
 In general, the (FUV-NUV) radial colour profiles become redder with  galactocentric
 distance in both rejuvenated ($\leq 4$ Gyr) and old ETGs. 
 Colour profiles of  
  NGC~1533, NGC~2962, NGC~2974, NGC~3489, and  IC~5063 show 
 rings and/or arm-like structures, bluer than the body of the galaxy,   
 suggesting the presence of recent star formation.
 Although seven of our ETGs show shell  systems in their optical image, 
 only NGC~7135  displays shells in the UV bands.    
 We characterize the UV and optical
 surface brightness profiles, along the major axis, using a Sersic law.
 The Sersic law exponent, $n$, varies from 1 to 16 in the UV bands. 
S0 galaxies tend to have lower values of $n$ ($\leq5$).
 The Sersic law exponent $n=4$ seems
 to be a watershed: ETGs with $n>4$ tend to have  [$\alpha$/Fe] greater than 0.15, implying a short 
 star-formation time scale.
 We find a significant correlation between the FUV$-$NUV colour and central velocity dispersions
$\sigma$, with the UV colours getting bluer at larger $\sigma$.
  This trend is likely driven by a combined effect of `downsizing'
  and of the mass-metallicity relation.   
 
\end{abstract}

\begin{keywords}
   {Galaxies: Galaxies: elliptical and lenticular, cD -- Galaxies: photometry -- 
 Galaxies: fundamental parameters --  Galaxies: formation -- Galaxies: evolution}
\end{keywords}

\section{Introduction}
The importance of ETGs  in the context of galaxy
formation rests on their evolved nature that makes them the 
fossil evidence of the process of galaxy evolution. Recent studies at
high redshift suggest the existence of a class of objects with
very similar morphological, dynamical and stellar population properties
to `bona fide', nearby ETGs \citep[see e.g.][]{Franx03,Chapman04,Treu05}. 
This raises the issue of how significant  galaxy   re-processing 
has been during the Hubble time, in particular at low redshift 
(i.e. in the range 0$\leq z \leq$1) where $\Lambda$CDM hierarchical 
models predict that most of the ETG stellar mass was assembled 
\citep{Delucia06}.

The study of the stellar populations in nearby galaxies
allows us to trace back with cosmic time the evolutionary
history of galaxies. Some recent studies paint a surprising 
picture for the evolution of ETGs showing that the environment
plays a key role in their evolution.
From the study of a sample of 124 galaxies, \citet{Thomas05}
find that massive ETGs in low density environments (LDEs) are on
average $\sim$2 Gyr younger and slightly ($\sim$0.05-0.1 dex) more metal
rich than their counterparts in high density environments.
Similarly, the study of \citet{Clemens09}, based on a sample of $\approx$14000
ETGs from the Sloan Digital Sky Survey (SDSS), suggests that  ETGs in dense environments are $\sim$20\%
older than those in LDEs. 
On the other hand, the environment does not affect  
$\alpha$-elements enhancement.  These results, obtained from the 
analysis of optical spectra, suggest that the evolutionary 
process is similar in the cluster and in the field but occurred earlier in 
more dense environments. ETGs in LDE frequently show 
signatures of relatively recent evolution, as indicated by the
presence of disturbed morphologies \citep[see e.g.][]{Reduzzi96,Colbert01}
 and of  kinematic sub-components \citep{McDermid06}.

\begin{table*}
\scriptsize{
\caption{The sample overview} 
\begin{tabular}{llcccccccccc}
\hline
{Ident.} & {RC3} &  {V$_{hel}$}      &  Dist.       & {r$_e$}    & $\epsilon$   & $\sigma_{r_e/8}$ & Age  & Z & [$\alpha$/Fe] & Activity  & $\rho_{xyz}$ \\
               &      &  {[km~s$^{-1}$]} &  {[Mpc]}&{[arcsec]}  &  &  {[km~s$^{-1}$]}     & {[Gyr]}  &    &       &Class  & [Gal. Mpc$^{-3}$]\\
\hline
& & & & & & & & & & \\
NGC~128  & S0 pec sp   & 4227 &   56.4  & 17.3 & 0.67 & 183 & 9.7$\pm$ 1.7 & 0.024 $\pm$ 0.004  & 0.16 $\pm$ 0.03 & LIN & \\
NGC~777  & E1          & 5040 &   67.2   & 34.4 &0.21 & 317 &    5.4     $\pm$       2.1   &      0.045   $\pm$      0.020     &    0.28    $\pm$     0.10   &  Sy/LIN& \\
NGC~1052 & E4         & 1475    & 17.8   & 33.7 & 0.28 & 215 &   14.5     $\pm$       4.2   &      0.032   $\pm$      0.007     &    0.34    $\pm$     0.05  & LIN & 0.49\\
NGC~1209 & E6:        & 2619   &   32.9  & 18.5 &0.52  & 240 &    4.8     $\pm$       0.9   &      0.051   $\pm$      0.012     &    0.14    $\pm$     0.02   & LIN & 0.13\\
NGC~1380 & SA0     & 1844    & 16.9    & 20.3 & 0.41  & 240   & 4.4 $\pm$ 0.7 & 0.038 $\pm$ 0.006 & 0.24 $\pm$ 0.02 & LIN & 1.54\\
NGC~1389 & SAB(s)0-: &  986& 16.9   & 15.0 & 0.37  & 139 &    4.5     $\pm$       0.6   &      0.032   $\pm$      0.005     &    0.08    $\pm$     0.02 &IN & 1.50\\
NGC~1407 & E0      & 1766  &  21.6  & 70.3 & 0.07 & 286 &    8.8     $\pm$       1.5   &      0.033   $\pm$      0.005     &    0.32    $\pm$     0.03 &IN & 0.42\\
NGC~1426 & E4     & 1443  &   16.8    & 25.0 & 0.34 & 162 &    9.0     $\pm$       2.5   &      0.024   $\pm$      0.005     &    0.07    $\pm$     0.05  &IN & 0.66\\
NGC~1453 &    E2  & 3906 &   52.1    &    25.0   & 0.17   & 289 &    9.1     $\pm$       2.8   &      0.034   $\pm$      0.009     &    0.22    $\pm$     0.05  &LIN & \\
NGC~1521 &    E3   & 4165 & 55.5    &    25.5  &0.35  &  235 &  3.2 $\pm$ 0.4 & 0.037 $\pm$ 0.006 & 0.09 $\pm$ 0.02 &LIN & \\
NGC~1533  & SB0-  &  773 &  13.4       & 30.0 & 0.19  & 174 &   11.9     $\pm$       6.9   &      0.023   $\pm$      0.020     &    0.21    $\pm$     0.10  &LIN & 0.89\\
NGC~1553 & SA(r)0  & 1280& 13.4     & 65.6  & 0.38 & 180 &    4.8     $\pm$       0.7   &      0.031   $\pm$      0.004     &    0.10    $\pm$     0.02  &LIN & 0.97\\
NGC~2911 & SA(s)0: pec  & 3131 & 41.7  & 50.9 & 0.32 & 235  & 5.7 $\pm$ 2.0 & 0.034 $\pm$ 0.019 & 0.25 $\pm$ 0.10 & LIN & \\
NGC~2962 & RSAB(rs)0+ & 2117 & 30.6 & 23.3 & 0.37 & && & & & 0.15\\
NGC~2974 & E4      & 1890   & 28.5    & 24.4 & 0.37 & 220 &   13.9     $\pm$       3.6   &      0.021   $\pm$      0.005     &    0.23    $\pm$     0.06 & LIN & 0.26\\
NGC~3258 & E1    & 2778 & 37.5     & 30.0& 0.13 & 271  &    4.5     $\pm$       0.8   &      0.047   $\pm$      0.013     &    0.21    $\pm$     0.03 & Comp. & 0.72 \\
NGC~3268 & E2          & 2818 & 37.5   & 36.1 & 0.24  & 227  &    9.8     $\pm$       1.7   &      0.023   $\pm$      0.004     &    0.34    $\pm$     0.04  & LIN & 0.69 \\
NGC~3489 & SAB(rs)0+  &  693 & 6.4  & 20.3  & 0.37 & 129 &    1.7     $\pm$       0.1   &      0.034   $\pm$      0.004     &    0.05    $\pm$     0.02 & Sy/LIN & 0.39\\
NGC~3607 & SA(s)0:    &  934 & 19.9  & 43.4 &0.11  & 220 &    3.1     $\pm$       0.5   &      0.047   $\pm$      0.012     &    0.24    $\pm$     0.03 & LIN  & 0.34\\
NGC~3818  & E5      &1701 & 25.1 & 22.2  & 0.36 &  191 &    8.8     $\pm$       1.2   &      0.024   $\pm$      0.003     &    0.25    $\pm$     0.03  &          & 0.20 \\
NGC~3962 & E1       & 1822 & 28.0 & 35.2&0.22 & 225 &   10.0     $\pm$       1.2   &      0.024   $\pm$      0.003     &    0.22    $\pm$     0.03 & LIN & 0.32 \\
NGC~4374 & E1      &1060  & 16.8 & 50.9 & 0.13 &  282 &    9.8     $\pm$       3.4   &      0.025   $\pm$      0.010     &    0.24    $\pm$     0.08 & LIN & 3.99\\
NGC~4552 & E           &  322 & 16.8 & 29.3 & 0.06 &  264 &    6.0     $\pm$       1.4   &      0.043   $\pm$      0.012     &    0.21    $\pm$     0.03 & Comp. & 2.97\\
NGC~4697 & E6      &1241 & 23.3 & 72.0  &0.32  & 174 &   10.0     $\pm$       1.4   &      0.016   $\pm$      0.002     &    0.14    $\pm$     0.04  & LIN & 0.60 \\
NGC~5011 & E1-2    &3104 & 40.9 & 23.8 & 0.15 & 249 &    7.2     $\pm$       1.9   &      0.025   $\pm$      0.008     &    0.25    $\pm$     0.06  & LIN & 0.27\\
NGC~5044 & E0      &2704 & 38.9 & 82.3 &0.11&  239 &   14.2     $\pm$       10.   &      0.015   $\pm$      0.022     &    0.34    $\pm$     0.17 & LIN & 0.38 \\
NGC~5363 & [S03(5)]       & 1138  & 22.4 & 36.1 & 0.34 & 199 &   12.1     $\pm$       2.3   &      0.020   $\pm$      0.004     &    0.16    $\pm$     0.05 & LIN &0.28  \\
NGC~5638 & E1   & 1676 & 28.4 & 28.0 & 0.11 & 165 &    9.1     $\pm$       2.3   &      0.024   $\pm$      0.008     &    0.24    $\pm$     0.05 & IN & 0.79\\
NGC~5813 & E1-2 & 1972 & 28.5  & 57.2 & 0.15 & 239 &   11.7     $\pm$       1.6   &      0.018   $\pm$      0.002     &    0.26    $\pm$     0.04 &LIN & 0.88 \\
NGC~5831 & E3   & 1656 & 28.5 & 25.5 & 0.15 & 164 &    8.8     $\pm$       3.5   &      0.016   $\pm$      0.011     &    0.21    $\pm$     0.09 & IN  & 0.83\\
NGC~5846 & E0+  & 1709 & 28.5 & 62.7 & 0.07 & 250 &    8.4     $\pm$       1.3   &      0.033   $\pm$      0.005     &    0.25    $\pm$     0.03 & LIN & 0.84\\
NGC~6868 & E2           & 2854 & 35.5 & 33.7 & 0.19 & 277 &    9.2     $\pm$       1.8   &      0.033   $\pm$      0.006     &    0.19    $\pm$     0.03 &LIN & 0.47 \\
NGC~6958 & E+         & 2652 & 35.4 & 19.8  & 0.15 &  223 & 3.0 $\pm$ 0.3 & 0.038 $\pm$ 0.006 & 0.20 $\pm$ 0.03 & Sy/LIN & 0.12\\
NGC~7079  & SB(s)0     & 2670 & 33.9  & 19.8 & 0.32 & 155 &    6.7     $\pm$       1.1   &      0.016   $\pm$      0.003     &    0.21    $\pm$     0.05    &LIN   & 0.19   \\
NGC~7135 & SA0- pec   & 2718 &34.7 &  31.4 & 0.31 & 231  & 2.2 $\pm$ 0.4 & 0.047 $\pm$ 0.010 & 0.46 $\pm$ 0.04 & LIN & 0.32\\
NGC~7192 & E+:         & 2904  & 35.6 &28.6 & 0.15 & 257 &    5.7     $\pm$       2.0   &      0.039   $\pm$      0.015     &    0.09    $\pm$     0.05  & LIN & 0.28  \\
NGC~7332 & S0 pec sp & 1207 & 18.2  & 14.7 & 0.42 &136 &    3.7     $\pm$       0.4   &      0.019   $\pm$      0.002     &    0.10    $\pm$     0.03 & &  0.12\\
IC~1459     & E          & 1659 & 20.0 & 34.4 & 0.28 & 311 &    8.0     $\pm$       2.2   &      0.042   $\pm$      0.009     &    0.25    $\pm$     0.04  &LIN & 0.28 \\
IC~4296    & E          & 3762 & 50.2 & 41.4 & 0.17 & 340 &    5.2     $\pm$       1.0   &      0.044   $\pm$      0.008     &    0.25    $\pm$     0.02 &LIN & \\
IC~5063    & SA(s)0+: & 3402 & 45.4 & 26.7 & 0.28  & 160 & & & &Sy &\\
& & & & &  & & & & & \\
\hline\hline
\end{tabular}}

Notes: See text for a detailed explanation of single columns. 
The age, metallicity and the $\alpha$-enhancement obtained from the Lick line-strength 
indices analysis, are obtained from Paper~III. The activity class is discussed 
in Paper~IV. We  used here the following
notation: LlN = LINER; Sy = Syfert like emission; Comp. = spectra can be due to either 
1) a combination of star formation and a Seyfert nucleus, or 2) a combination of 
star formation  and LINER emission \citep{kewley06}, 
IN = either faint or no emission lines. 
 
\label{tab1}
\end{table*}

The advent of the  Galaxy Ultraviolet Explorer ({\it GALEX}, \citet{Martin05}) has 
contributed to strengthen the idea that recent star formation is often present
in ETGs. Through {\it GALEX} ultraviolet imaging, 
\citet{Schawinski07} found  
that 30\% of massive ETGs show ongoing star formation and that this fraction 
is higher in LDEs. \citet{Rogers07} concluded that ETGs in 
low density environments are less likely to present weak episodes 
of recent star formation than their high-density counterparts.
{\it GALEX} studies of ETGs with shell structures 
are of particular relevance, since these features, believed to be  `bona fide' 
signatures of recent accretion/merging events, characterize a significant fraction
 ($\approx$16.5\%) of field ETGs \citep{MC83}. Combined analyses of
Lick line-strength indices (e.g. Mg2, H$_\beta$, H$_{\gamma A}$, H$_{\delta A}$) 
and UV colours, revealed the presence of recent star formation
in shell galaxies \citep{Rampazzo07,Marino09}. 
Similar results have been recently obtained by \citet{Jeong09} for a {\tt SAURON} 
galaxy sample. They determined   
the UV Fundamental Plane and
suggested that the dominant fraction of the tilt and the scatter of this plane
 is due to the presence of young stars in preferentially low mass ETGs. 

This is the fifth paper of a series dedicated to the study of ETGs with emission lines 
(\cite{Rampazzo05} hereafter Paper~I; \cite{Annibali06}
hereafter Paper~II; \cite{Annibal07} hereafter Paper~III;
\cite{Annibali10} hereafter Paper~IV).
Our program considers 65 ETGs  mainly located in groups and
low density environments (LDEs). 
In paper III, we showed that the ETGs of our sample have a large age spread,
with luminosity-weighted ages ranging from ~1 Gyr to a Hubble time.
Luminosity-weighted ages younger than ~5 Gyr are always associated
with the galaxies residing in the lowest density environments.
We suggest that such young ages are due to `rejuvenation' episodes,
which do not involve more than 25\% of the total galaxy stellar mass.
In Paper IV we investigated the nature of the ionization mechanism
through standard emission-line diagnostic diagrams
(e.g. log [OIII]/H$\beta$ vs. log [NII]/H$\alpha$).
The majority of our ETGs are classified as LINERs.
Galaxies classified as Seyferts in our sample tend to have
very young ($<$5 Gyr) mean ages, supporting that
star formation and  AGN phenomena co-exist
\citep[see e.g.][]{Terlevich90,Wada04}.
The gas oxygen abundance, derived through the emission lines,
gives systematically lower values than that of stars.
A possible explanation is that the gas has an external origin,
e.g. derived from the accretion of a metal poorer companion.

We present in this paper GALEX FUV and NUV surface photometry of 40 
ETGs.  
The UV wavelengths are particularly sensitive to young populations,
and thus ideal to characterize any recent star formation. 
The paper is organized as  follows.
In Section 2 we describe the sample. Sections 3 presents the 
{\it GALEX} UV observations, the data reduction
and the  comparison of our results with the 
 literature. In Section~4 we present our results. 
We discuss the galaxy morphological peculiarities,
the shape of the surface brightness profiles,
the UV radial colour profiles.
In Appendix A we summarize, for the individual galaxies, 
the UV and optical photometric properties,
as well as the kinematic and morphological peculiarities.
In Appendix B we include the $r$-band surface photometry
obtained for a subset of 14 galaxies from SDSS data.
 
\section{Sample}

A sample of 40 galaxies is selected from a wider
sample of 65 ETGs (see Paper~I and II).  
Table~\ref{tab1} reports the main properties of the sample.
Column (1) gives the galaxy identification; 
column (2) provides the galaxy morphological classification
according to RC3 \citep{RC3}.
Only in a few cases  does the RC3 differ from that of the 
RSA \citep{RSA}: NGC~4552, NGC~5846, NGC~6958 and 
NGC~7192 are classified as S0 in RSA and as E in RC3.  
Column (3) gives the galaxy systemic velocity, V$_{hel}$, which is lower
than $\sim$5000 km~s$^{-1}$ in all cases. For galaxies with a heliocentric 
systemic velocity  lower than 3000 km~s$^{-1}$ we adopt the distances 
(column 4) provided by  \citet{Tu88}.  For larger velocities, distances 
assume a Hubble  Constant of  H$_0$=75 km~s$^{-1}$Mpc$^{-1}$. 
Columns (5) and (6) provide  the optical effective radius, r$_e$, i.e the radius of the circular
aperture encircling half of the total galaxy optical light, and the average ellipticity, 
as obtained from RC3  and {\tt HYPERCAT}.
In column (7) we report the galaxy central velocity dispersion, 
within an aperture of 1/8 of the effective radius. 
\citet{Cappellari06} has investigated the relation
between galaxy mass-to-light ratio and the line-of-sight
component of the velocity dispersion. 
They provide relations (their equation 7 and 10)
which allow a direct transformation from the measured 
velocity dispersion to galaxy mass. According 
to the above formulae, the quoted velocity dispersion, 
that ranges from $\approx$130 to 340 km~s$^{-1}$, 
indicate that our galaxies cover a large range in masses (roughly 3 $\times$ 10$^{10}$ - 9 $\times$ 10$^{11}$ 
M$_{\odot}$). 
 
The luminosity-weighted ages as derived in Paper~III are given in column~(8);
they range from 1.7$\pm$0.1 Gyr  (NGC~3489) to 14.5$\pm$4.2 Gyr (NGC~1052). 
Elliptical galaxies are on average older than lenticulars. 
A definite trend of increasing metallicity (column 9) and
[$\alpha$/Fe] (column 10) with the velocity dispersion is observed, testifying that 
the chemical enrichment was more efficient and the duration of the star
formation shorter in more massive galaxies. These  two relations do 
not depend on the galaxy morphological type.

 Nebular emission lines are commonly found in
the inner regions of ETGs: in the optical, the fractions of documented detection
in different samples are 55\%-60\% \citep{Phi86}, 
72\% (ellipticals, E) -85\% (lenticulars, S0)
\citep{Mac96}, 66\%(E) -83\% (S0) \citep{Sarzi06}, 
and 52\% \citep{yan06}. Concerning a possible environmental
effect on the presence of emission lines in ETGs, \citet{Grossi09}, in a 
sample of 62 ETGs in low density environments, found that 44\% of the luminous
ETGs (M$_B<-$17) are detected in HI (i.e. 10 times more frequently than in 
the Virgo cluster),  60\% of which show emission line ratios typical of 
star-forming galaxies.
Optical emission lines are detected in 89\% of our original sample of 65
ETGs (Paper~IV). For the  sample of 40 ETGs studied  
in this paper, the detection drops to 82\%. 
Incidence and strength of line emission do not correlate either with the 
E/S0 classification, or with the fast/slow rotator classification.

By means of optical emission line ratios (e.g., [OIII]$\lambda$5007/H$\beta$ and
[NII]$\lambda$6584/H$\alpha$, \citet{Baldwin81}), we have shown
that for the majority of our ETGs  the emission  is ``indistinguishable''
from that of low-ionization nuclear emission-line regions (LINERs, \citet{heck80})
in agreement with other studies of ETGs \citep[see e.g.][]{Phi86,Gou99,kewley06}.
The activity class, as derived in Paper~IV from optical emission line
diagnostic diagrams, is given in column~11 of Table 1. 

The richness of the environment in which our ETGs are located
 is well described by the  parameter $\rho_{xyz}$ (column 12 of Table 1) provided by
\citet{Tu88} in the {\it Nearby Galaxy Catalog}.  It represents the density of
galaxies brighter than B=$-$16 mag in the vicinity of each galaxy in
Mpc$^{-3}$. 
The galaxies of our sample are mainly located in low
density environments. The local density around our galaxies varies from
$\rho_{xyz}$ $\approx$ 0.1, characteristic of nearly isolated galaxies, to
$\rho_{xyz}$ $\approx$ 4, which is characteristic of dense galaxy
regions in the Virgo cluster. 
For comparison, in the \citet{Tu88} catalogue  NGC~1380
and NGC~1389, Fornax cluster members, have values of $\rho_{xyz}$=1.54
and 1.50 respectively.  Thus our sample, although biased towards low
density environments, contains a fraction  of galaxies in relatively
dense environments.

\section{Observations and Data Reduction}

The UV imaging was obtained with  GALEX  \citep{Martin05, Morrissey07}
 in two ultraviolet bands, FUV (1344 -- 1786 \AA) ~ and NUV (1771 -- 2831 \AA). 
The instrument has a very wide field of view (1.25 degrees diameter) and a spatial resolution  
$\approx$4\farcs2  and 5\farcs3 FWHM in FUV and NUV respectively, 
sampled with 1\farcs 5$\times$1\farcs 5 pixels \citep{Morrissey07}.

\begin{figure*} 
\begin{tabular}{cc}  
\vspace{-0.25cm}
 \psfig{figure=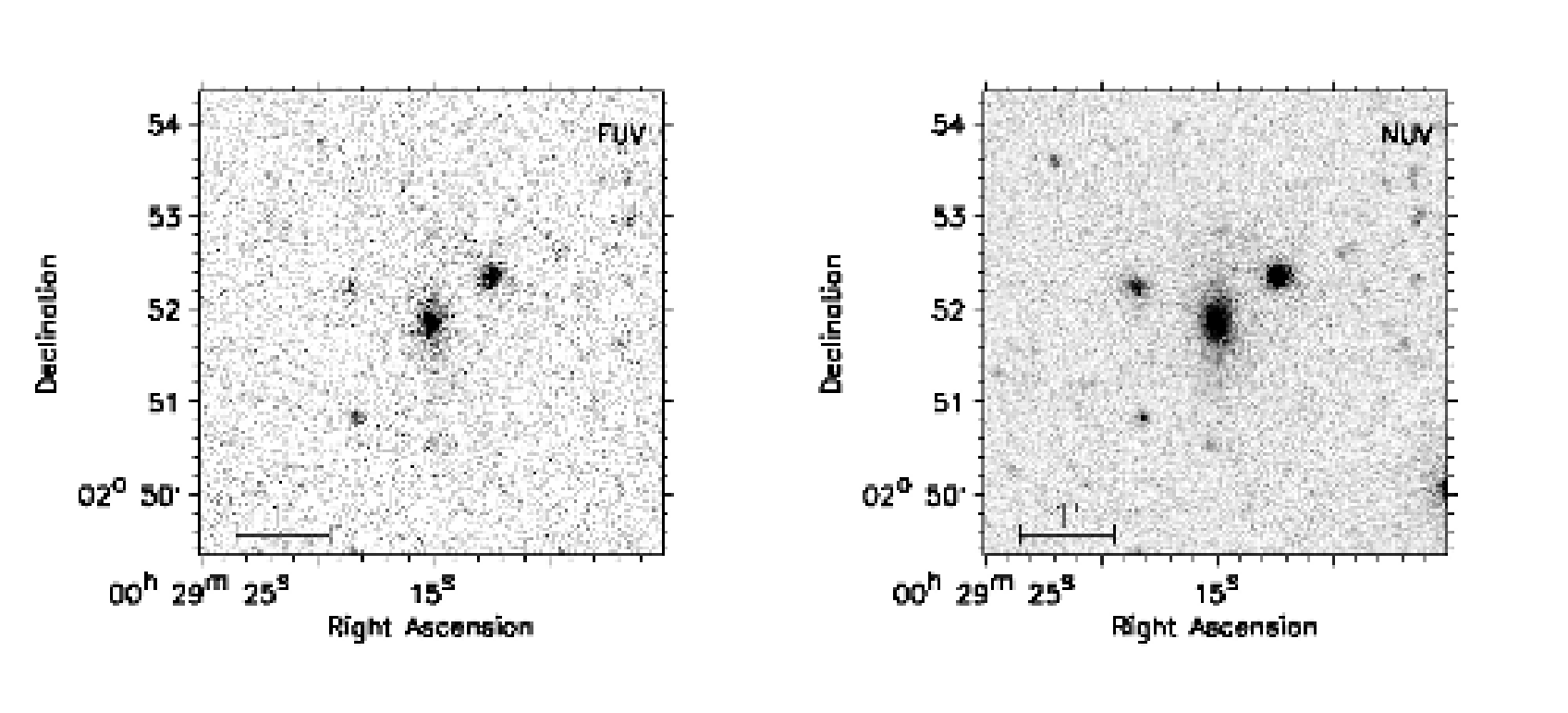,width=11cm} & 
 \vspace{-0.25cm}
 \psfig{figure=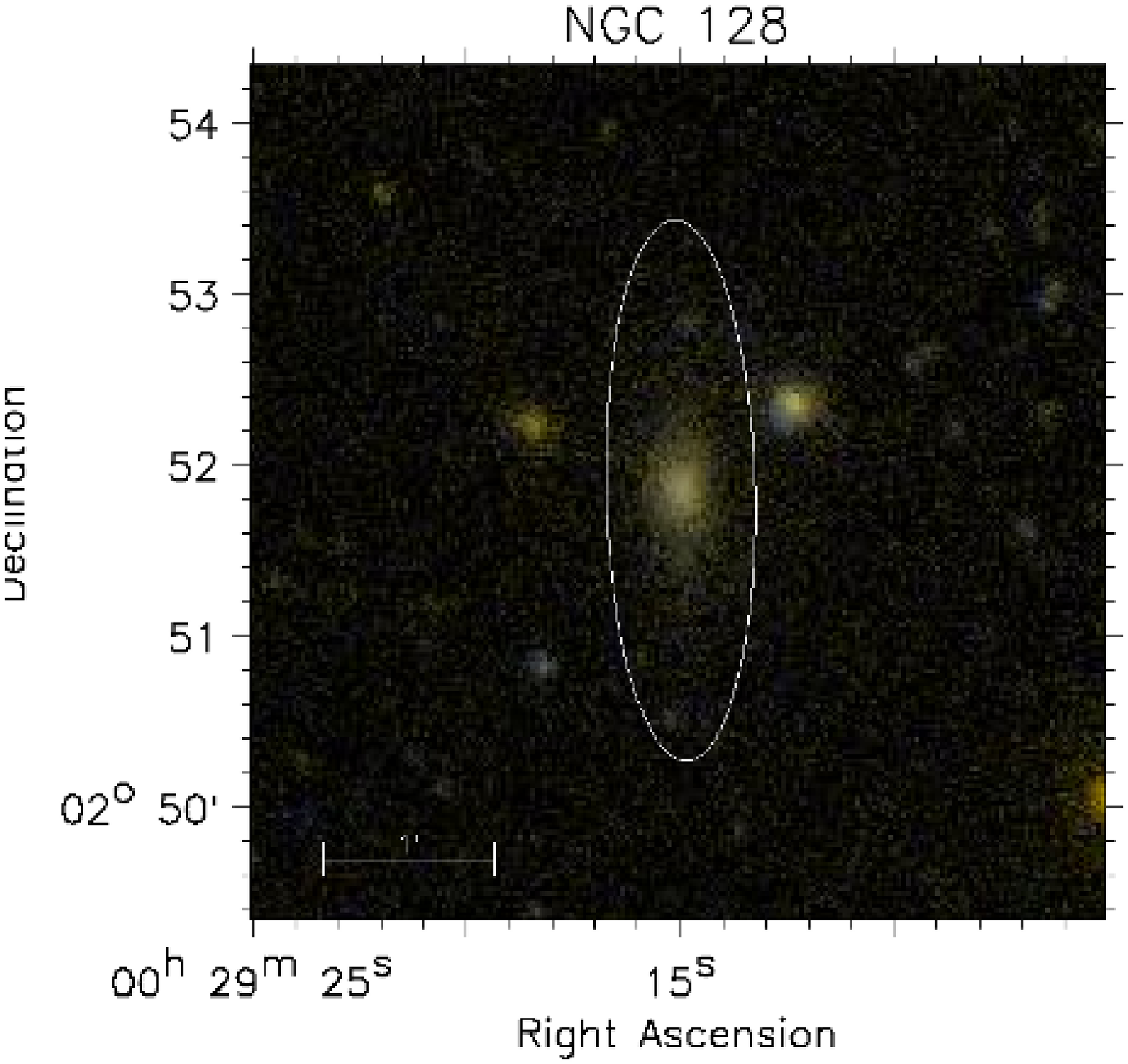,width=5cm} \\
\vspace{-0.25cm}
\psfig{figure=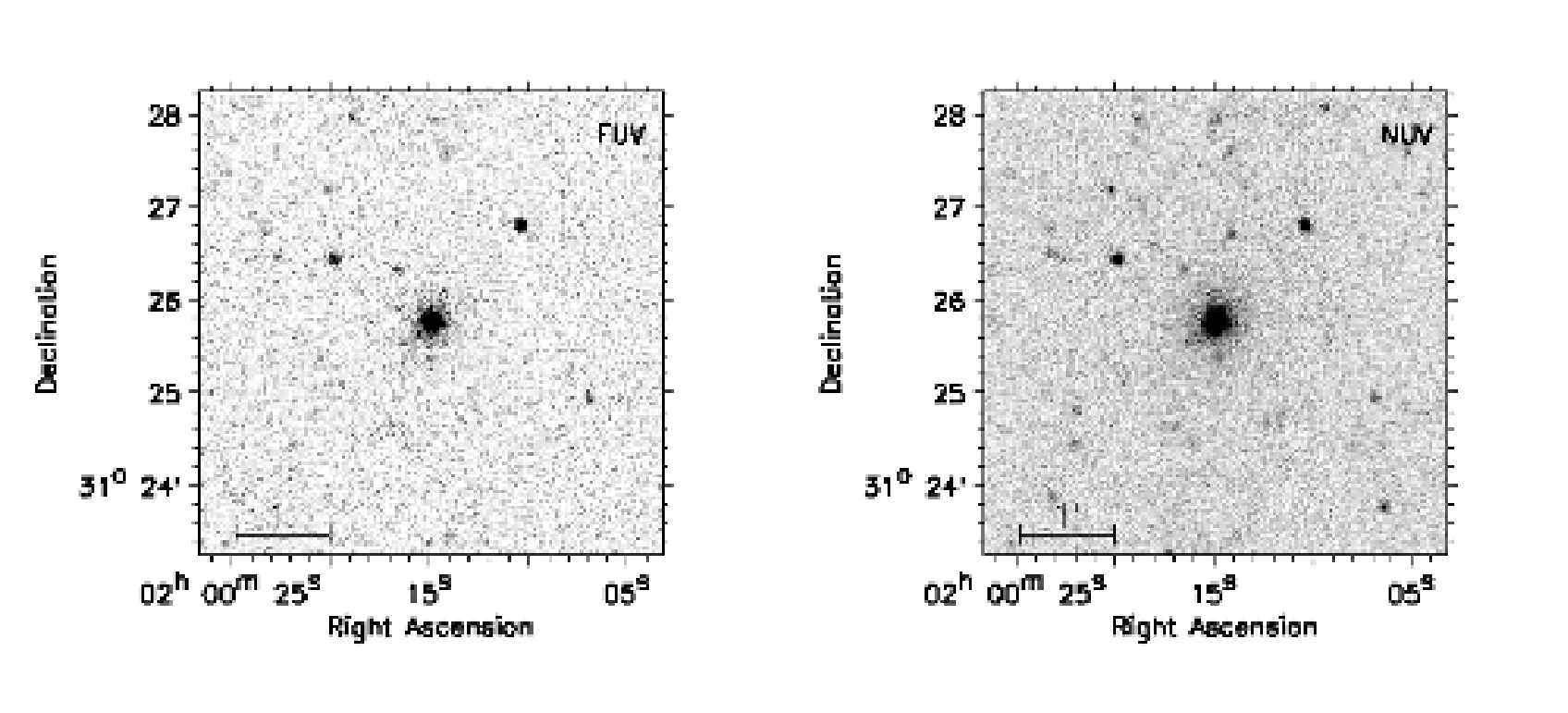,width=11cm} & 
\vspace{-0.25cm}
\psfig{figure=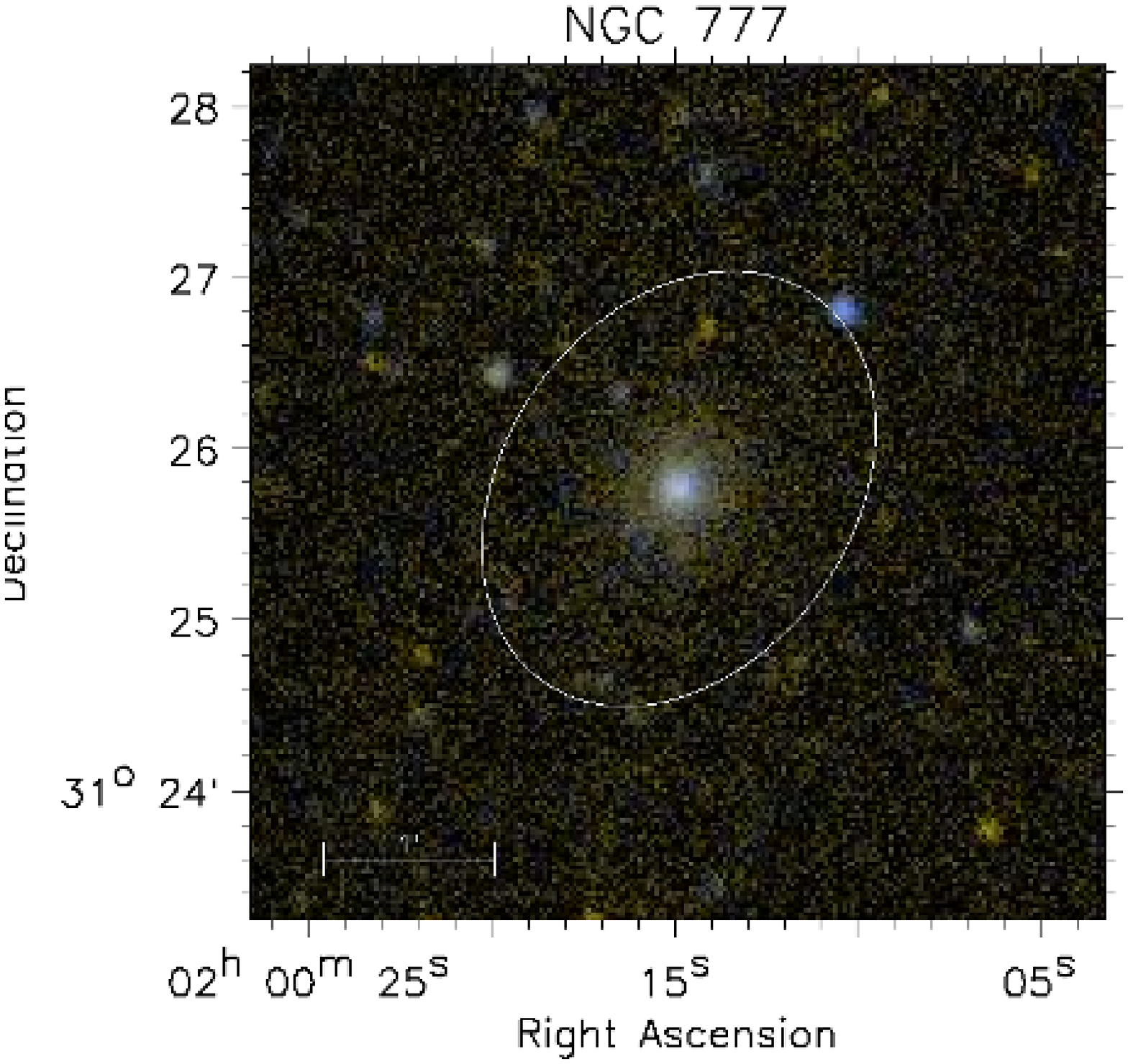,width=5cm} \\ 
\vspace{-0.25cm}
\psfig{figure=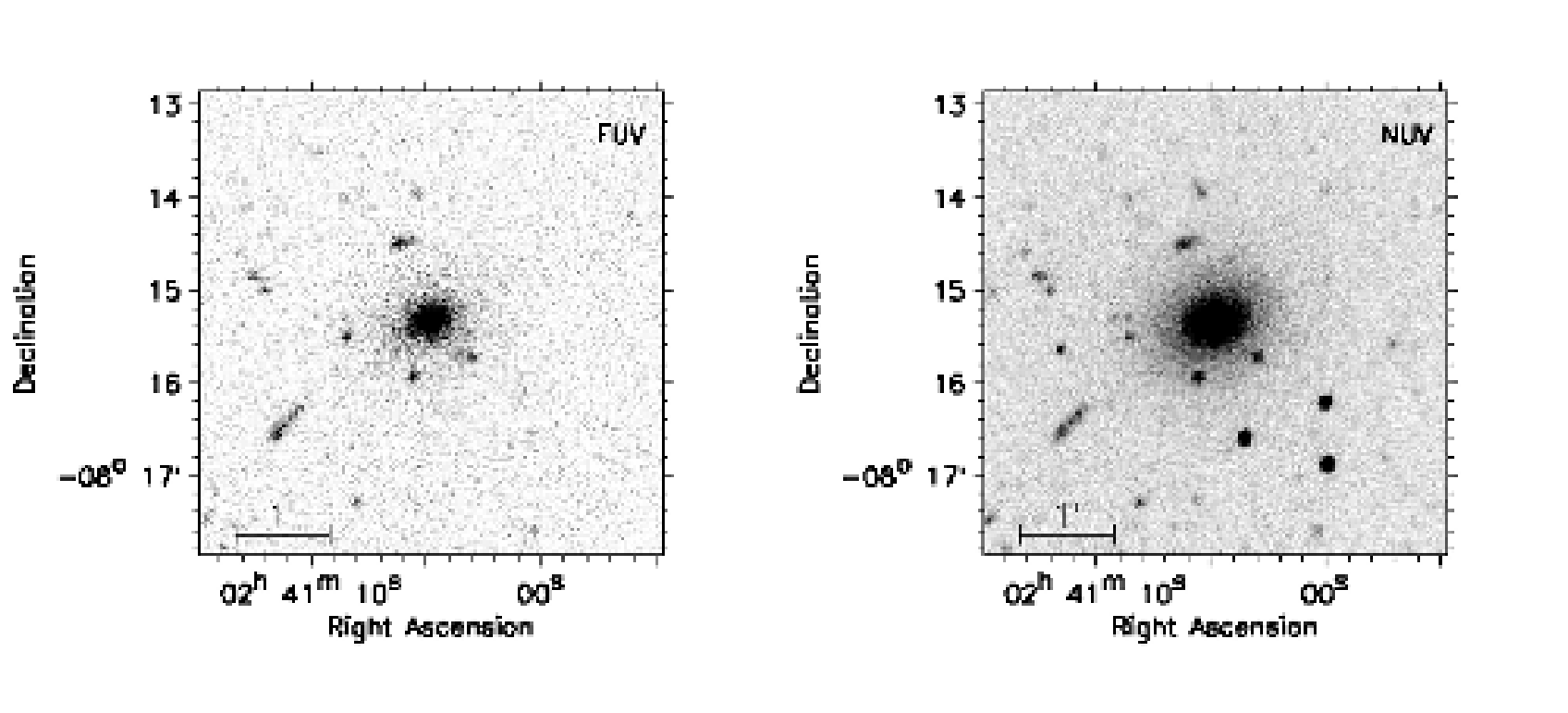,width=11cm} & 
\vspace{-0.25cm}
\psfig{figure=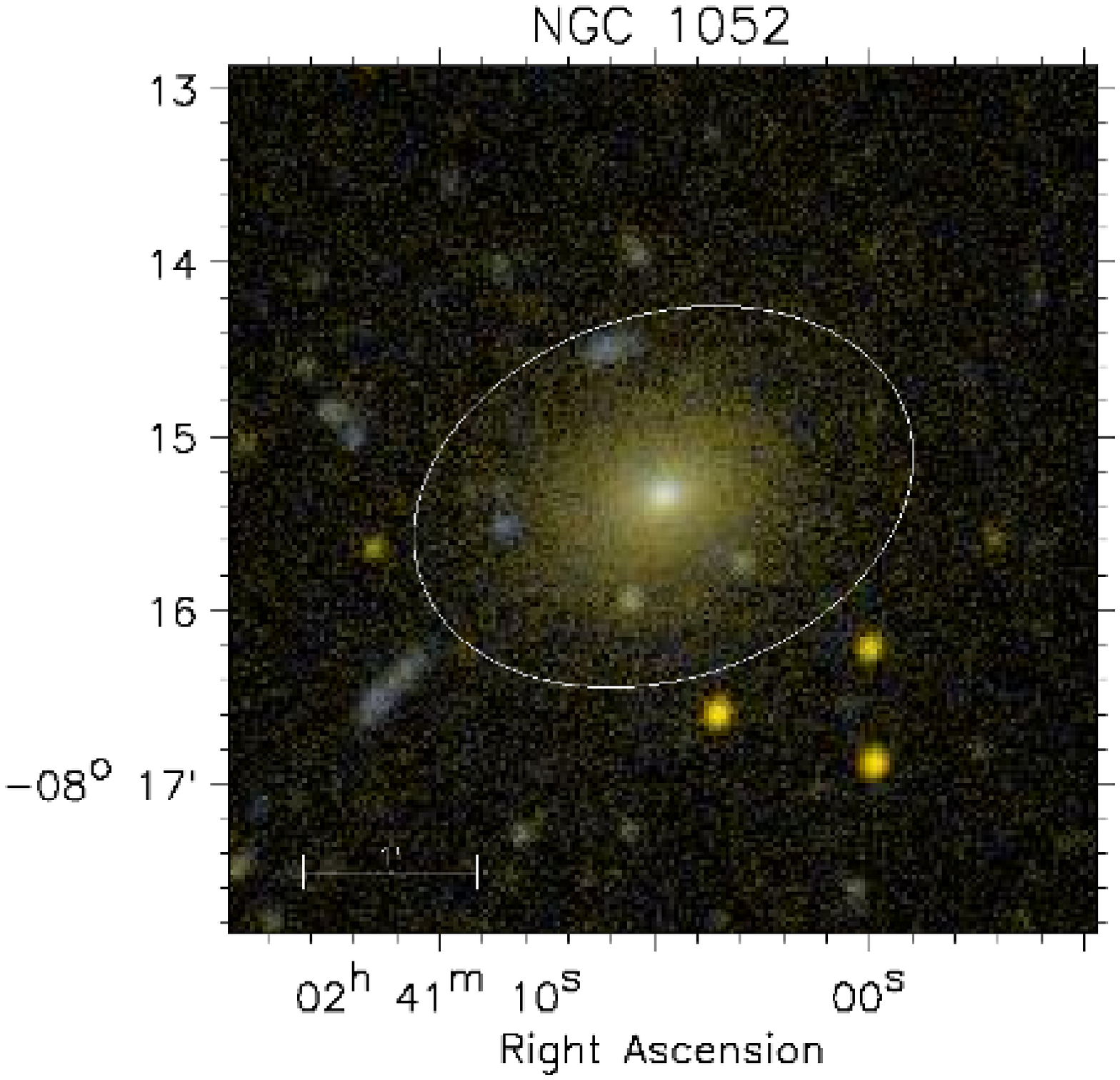,width=5cm}\\ 
\vspace{-0.25cm} 
\psfig{figure=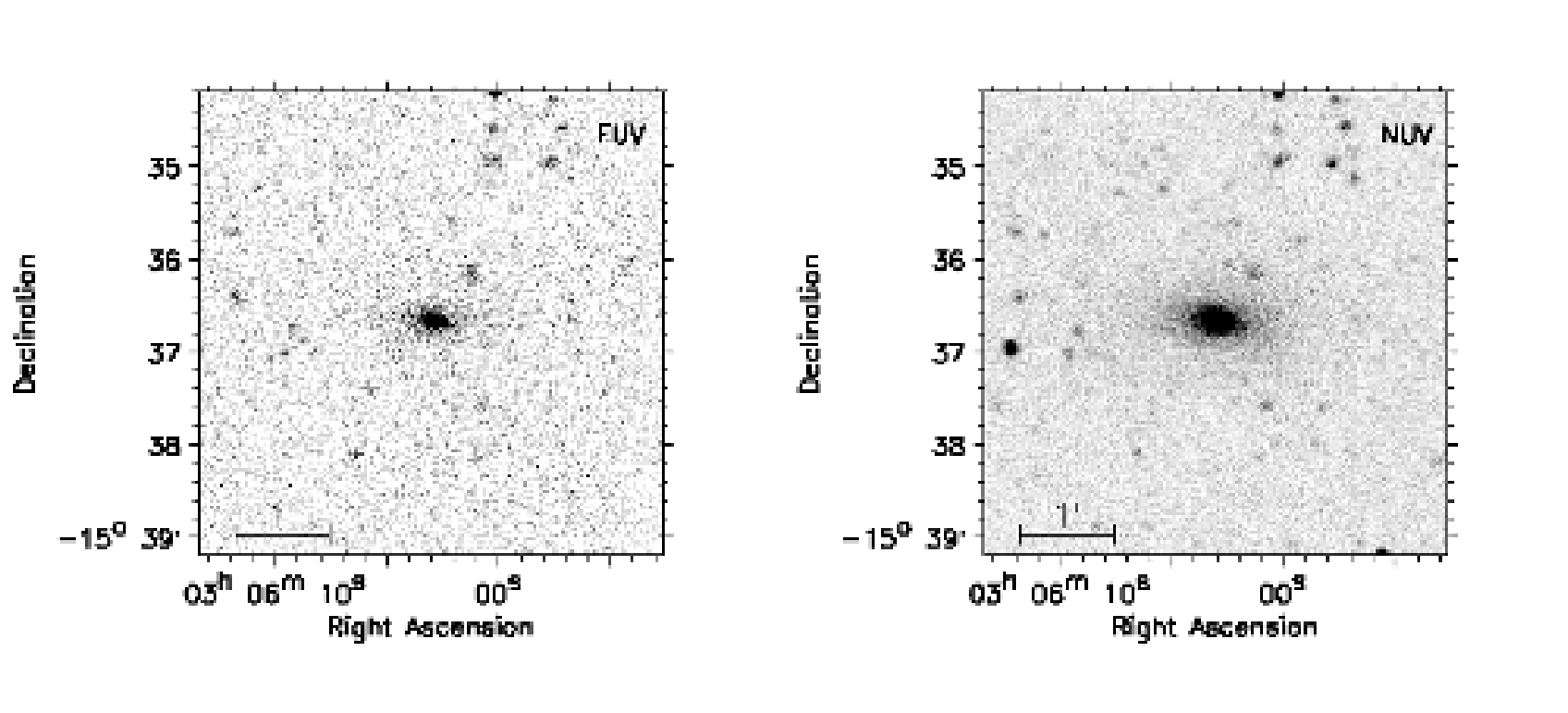,width=11cm} & 
\vspace{-0.25cm}
\psfig{figure=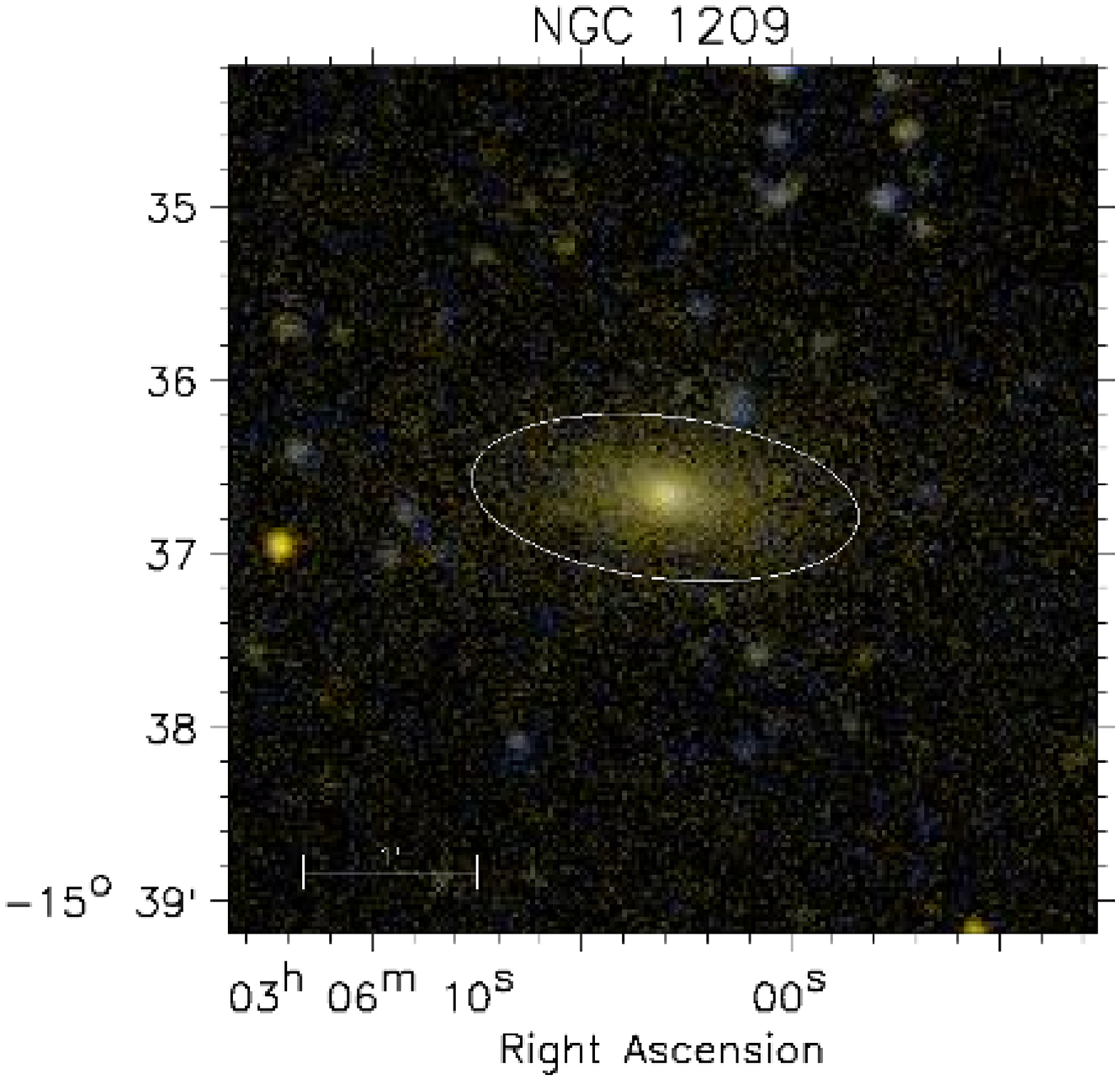,width=5cm} \\
\vspace{-0.25cm}
\psfig{figure=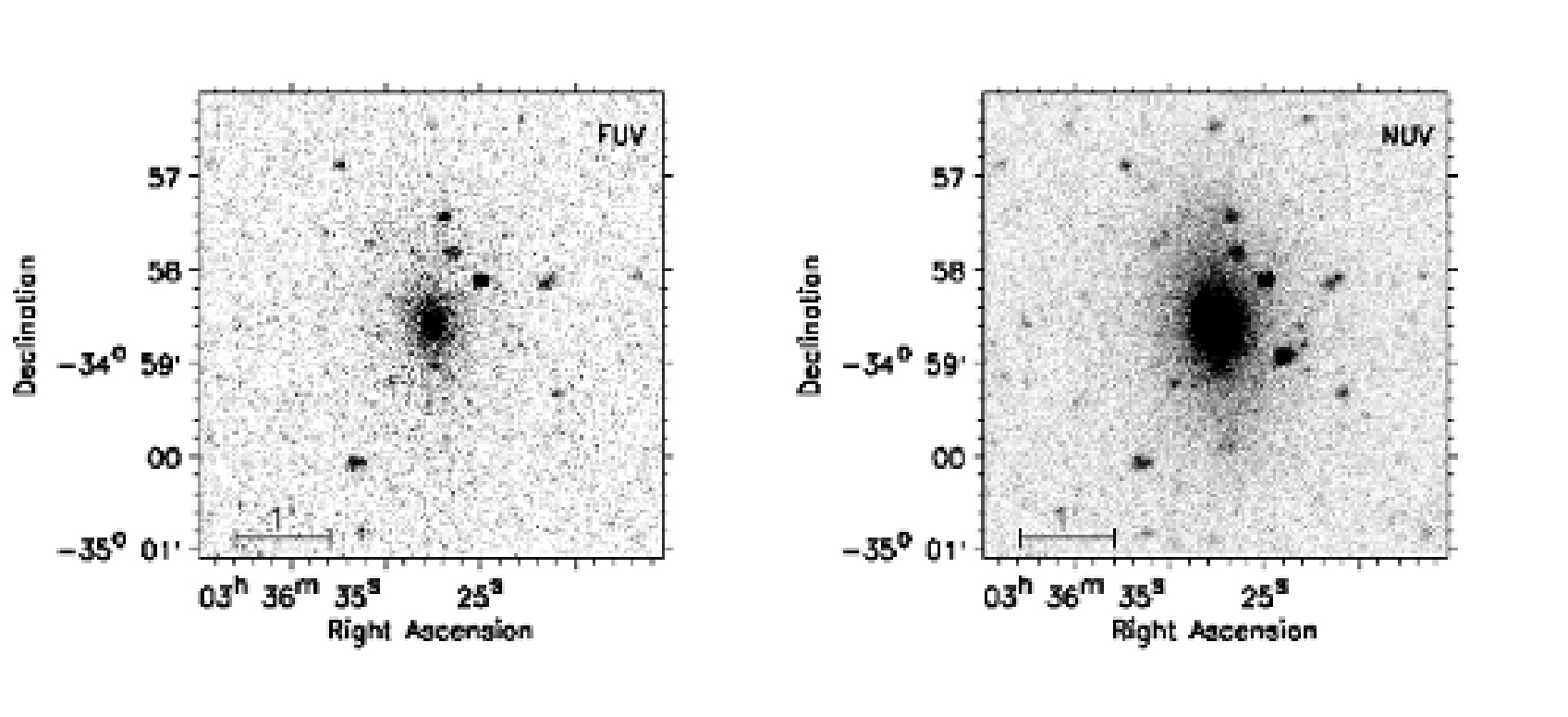,width=11cm} & 
\vspace{-0.25cm} 
 \psfig{figure=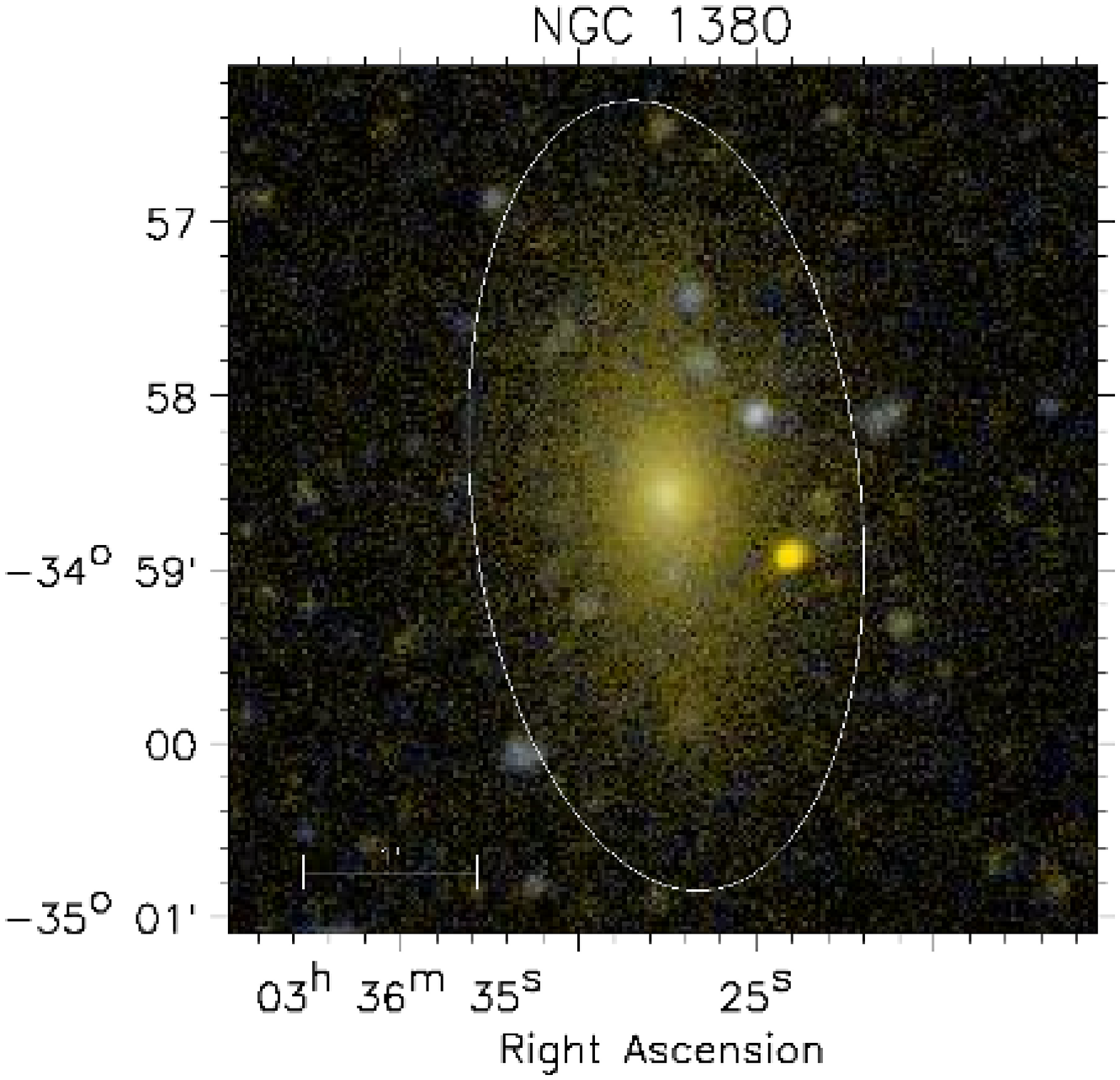,width=5cm}\\ 
  \end{tabular} 
    \caption{Left and middle panels: {\it GALEX} FUV and NUV background 
  subtracted images of the sample. Right panel: false colour images (FUV Blue; NUV yellow). 
Ellipses mark the optical D$_{25}$ diameter, within which we measure the 
NUV and FUV integrated magnitudes. Figure 1 is available in its entirety in the the online version of the Journal. 
A small portion is shown here for guidance.}
   \label{fig1}
\end{figure*}

The  sample we analyze is composed of 16 ETGs observed in our 
Cycle~3 proposal (ID GI3-0087 PI R. Rampazzo) and 24 ETGs taken
from the public {\it GALEX} archive (see Table 2). 
The exposure times for most of our sample are $\sim$ 1500 seconds
(limiting magnitude in FUV/NUV of $\sim$ 22.6/22.7 AB mag \citep{Bianchi09}. 
NGC 1389 and NGC 5363 have exposure times $\sim$10 times longer   
($\sim$2.5 mag fainter limit).
For each galaxy, in Figure~\ref{fig1} we show the {\it GALEX} FUV 
and NUV  images and a false colour composite image.   

Additional {\it GALEX}  observations with exposure times of 
about 100 seconds exist for the following sub-set of the  
Paper~I plus Paper~II samples: 
NGC~1947, NGC~5193, NGC~6721,
NGC~6758,  NGC~7007 and  NGC~7377 but were not considered 
in the present paper, in order to have an homogeneous flux limit.

\begin{figure*}
\psfig{figure=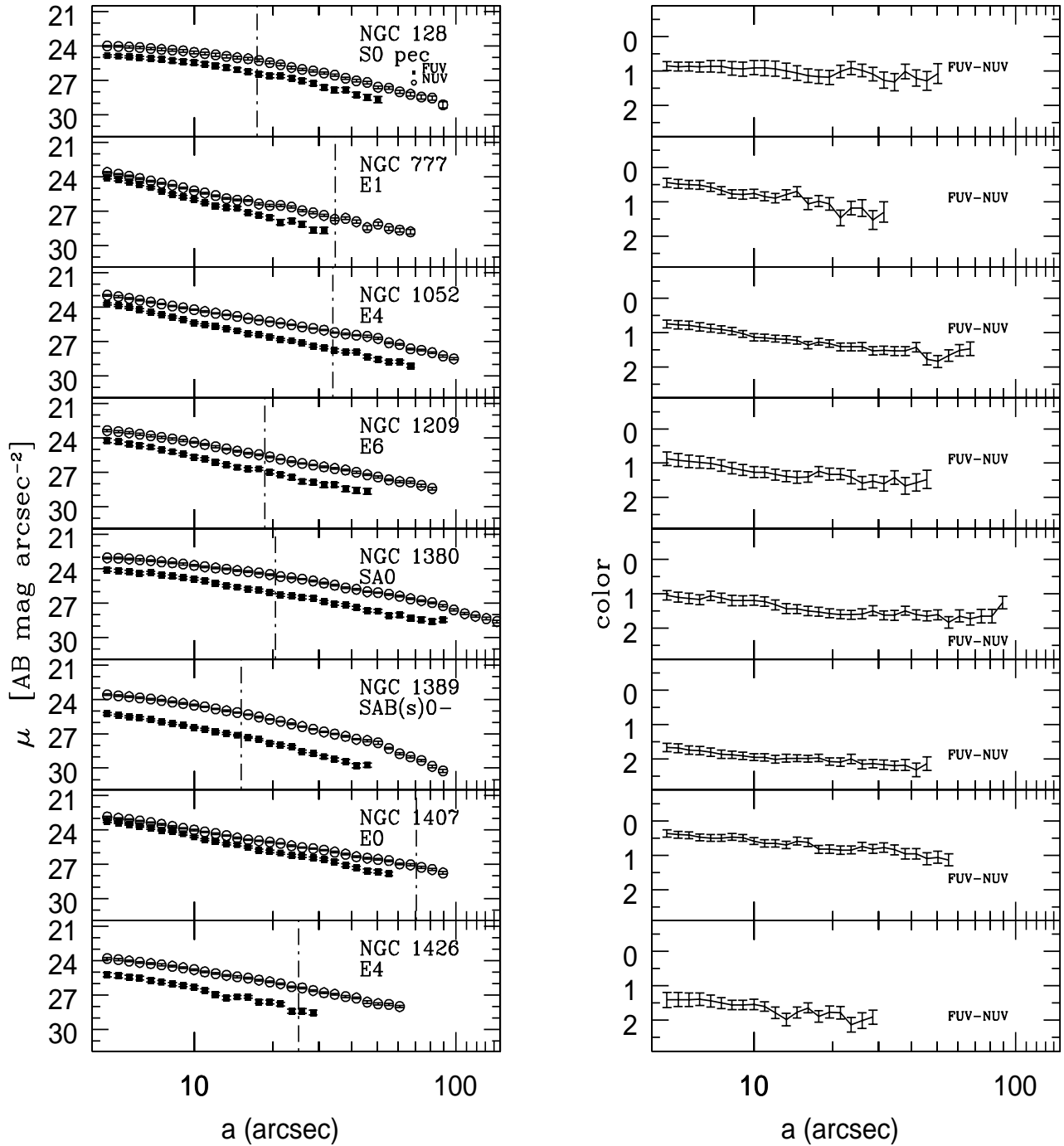,height=20.0cm,width=18.5cm} 
\caption{From top to bottom: (left panels) Luminosity profiles along the semi-major axis of the fitted 
ellipse in the {\it GALEX} FUV and NUV bands. The vertical dot-dashed lines indicate the optical effective 
radius (see Table~1).(right panel) Radial (FUV-NUV) colour profile v.s. semi-major axis. 
 }
   \label{fig2}
\end{figure*}

\begin{figure*}
\psfig{figure=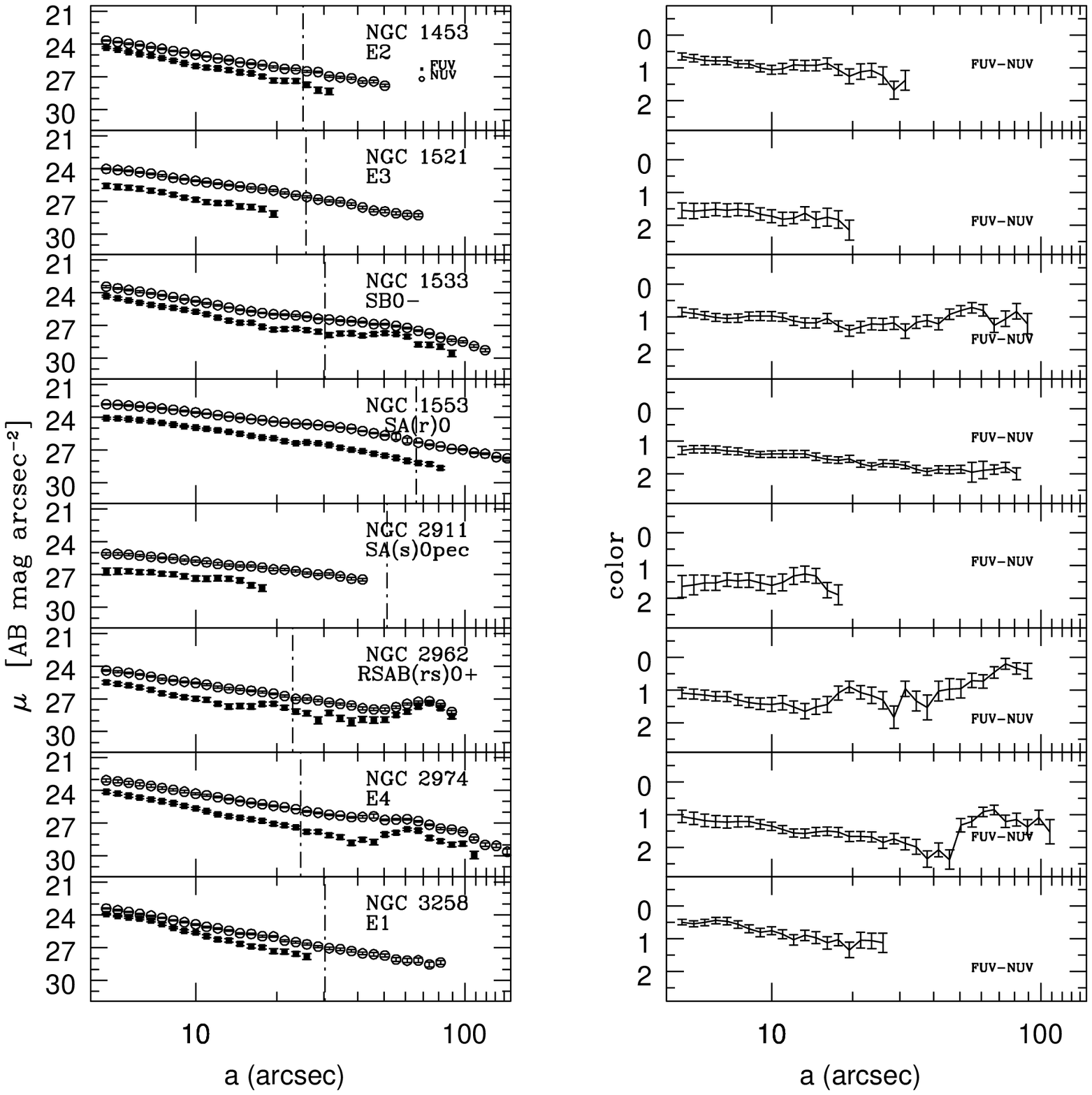,height=20.0cm,width=18.5cm} 
\addtocounter{figure}{-1}
\caption{Continued.} 
 \end{figure*}

\begin{figure*}
\psfig{figure=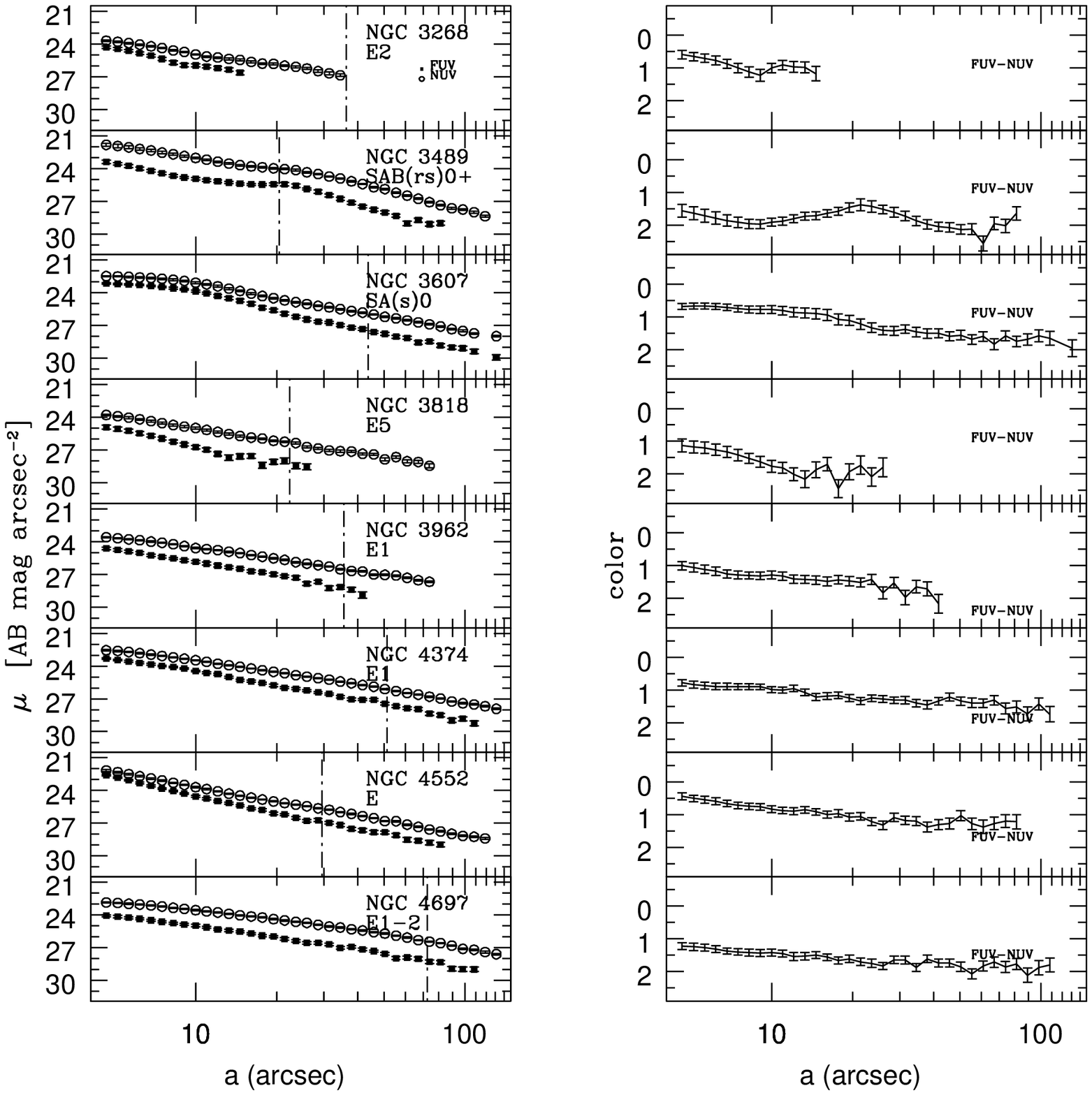,height=20.0cm,width=18.5cm} 
\addtocounter{figure}{-1}
\caption{Continued.} 
 \end{figure*}

\begin{figure*}
\psfig{figure=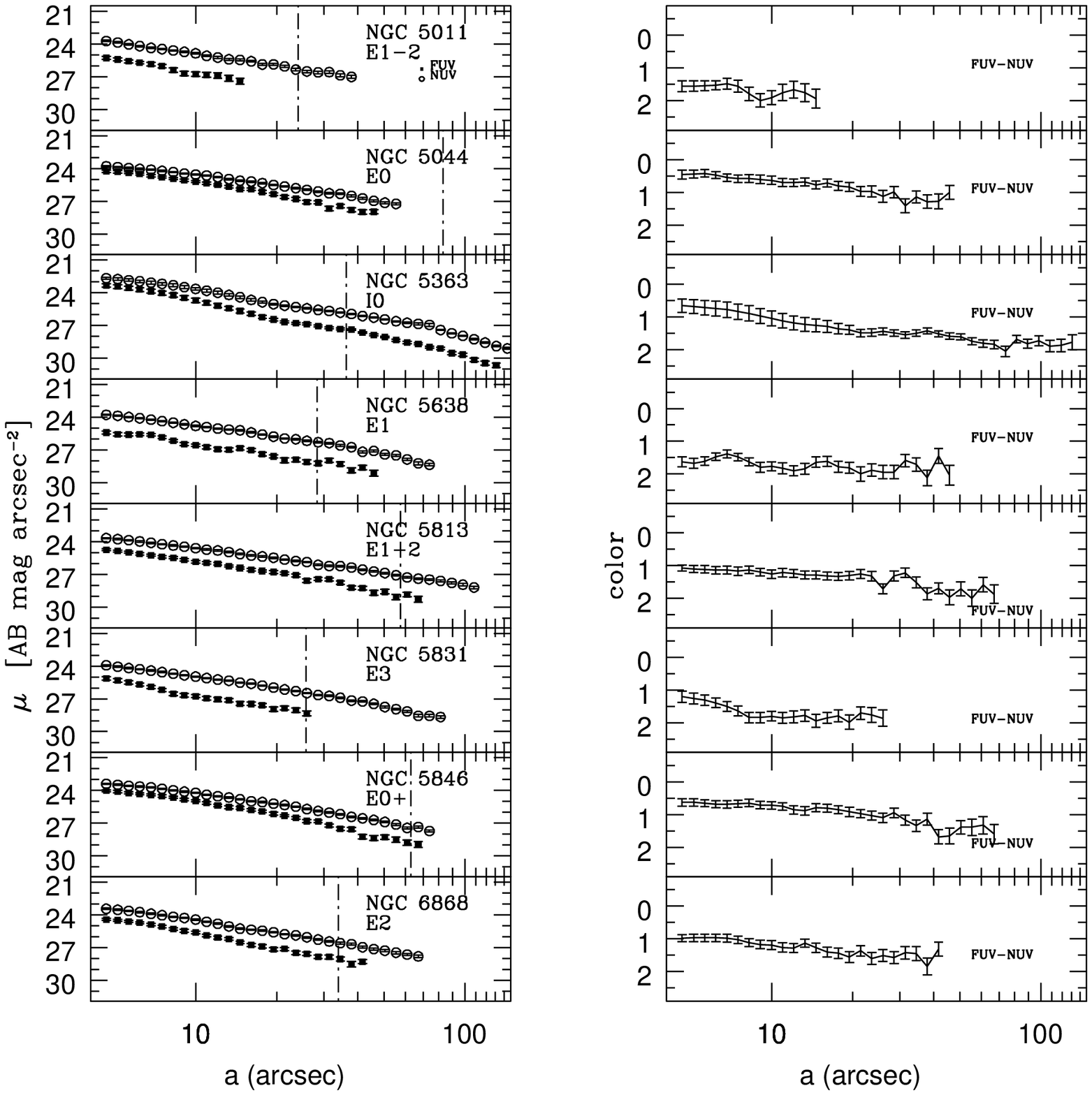,height=20.0cm,width=18.5cm} 
\addtocounter{figure}{-1}
\caption{Continued.} 
\end{figure*}

\begin{figure*}
\psfig{figure=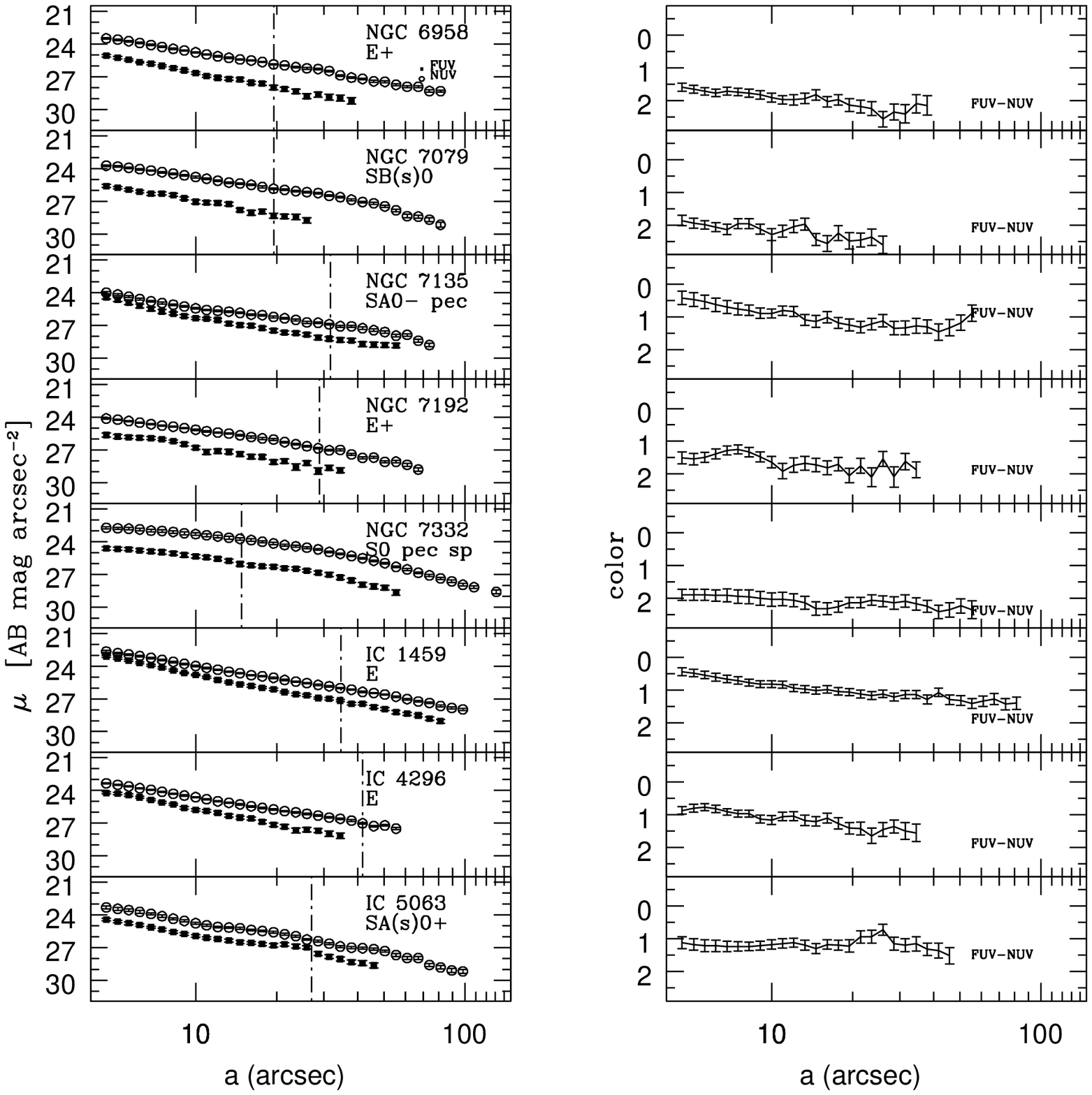,height=20.0cm,width=18.5cm} 
\addtocounter{figure}{-1}
\caption{Continued.} 
\end{figure*}
 

We used background-subtracted FUV and NUV {\it GALEX}  intensity 
images to perform  photometry. 
Table~3 lists the FUV and NUV magnitudes in the AB system of the 
ETG sample  within  r$_{e}$/4, r$_{e}$/8 and D$_{25}$ apertures, where r$_e$ and
D$_{25}$ are the optical effective radius (see Table~1) and the diameter 
of the isophote at $\mu_B$=25 mag~arcsec$^{-2}$ respectively (taken from {\tt HYPERLEDA}).  
The   D$_{25}$ ellipses are shown in Figure~\ref{fig1} . 
FUV and NUV magnitudes were computed
as m$_{UV}$ = -2.5 $\times$ log~CR$_{UV}$ +ZP,  where CR is the 
dead-time-corrected flat-fielded count rate, and the zero points ZP=18.82 and 20.08 mag in FUV
and NUV respectively \citep{Morrissey07}. 
FUV and NUV  magnitudes
and photometric errors were determined from the original un-smoothed images,
after manually subtracting foreground stars.
In order to estimate the errors on UV 
magnitudes, we propagated the Poisson statistical errors on source and background
counts.   Background counts were computed measuring
the background images provided by the GALEX pipeline in the same D25 aperture of the galaxy. 
In addition to the statistical error, we added an uncertainty to account
for systematic inaccuracies in the zero point of the absolute calibration of 
0.05 and 0.03 magnitudes for FUV and NUV respectively \citep{Morrissey07}.

The comparison between  our D$_{25}$ magnitudes and those
of \citet{Gil07} shows  general agreement.
Discrepant measurements are found only in the FUV for NGC~5813 
(D$_{25 our}$-D$_{25 GdP}$=0.42), NGC~777 (0.29) and  NGC~5638 (0.20).
For these objects, the D25 sizes adopted by \citet{Gil07}, taken from RC3 
slightly differ from ours, taken from {\tt Hyperleda}, and may account for
the discrepancies.

The surface photometry was carried out using the {\tt ELLIPSE} fitting routine
in the {\tt STSDAS} package of {\tt IRAF}.  
{\tt ELLIPSE} computes a Fourier expansion for each successive
isophote \citep{Jedrzejewski87}, resulting in  photometric
surface profiles.  
We masked the foreground objects in the regions where we evaluated the
surface brightness profiles. 
Surface photometry was corrected for galactic extinction assuming Milky Way dust with Rv=3.1 \citep{Cardelli89}, 
A$_{FUV}$/E(B-V)= 8.376 and A$_{NUV}$/E(B-V)= 8.741. 
The UV radial profiles are truncated when the uncertainty in the 
surface brightness exceeds 0.3 mag~arcsec$^{-2}$. The UV limiting surface 
brightness is  $\sim$ 28 AB mag/arcsec$^2$.   Radial (FUV-NUV) colour profiles 
generally become redder from the centre to the periphery 
of the galaxy. Significant deviation from this general behaviour is found in the 
galaxies showing ring and/or arm-like  features.  

In Figure \ref{fig2} we present the {\it GALEX} FUV and NUV surface brightness profiles (left panels) 
along the major axis and the (FUV-NUV) radial colour profiles.
A short description of individual galaxies is given  in the Appendix~A.

Six ETGs were also measured by \citet{Jeong09}. 
There are no systematic differences among their and our measures. 
The largest difference is found  for
NGC~2974: the average difference in the FUV and the NUV surface brightness
is $\langle (FUV_{Jeong09}-FUV_{our}) \rangle$=0.23$\pm$0.12  mag and 
$\langle (NUV_{Jeong09}-NUV_{our}) \rangle$=0.17$\pm$0.10  mag, likely due
to the estimate of the background, determined
following different approaches by the two studies. 
For the other galaxies the average differences are generally smaller, 
sometimes lower than formal errors. 
In particular: NGC~4374 $\langle(FUV_{Jeong09}-FUV_{our})\rangle$=-0.08$\pm$0.10 and 
$\langle (NUV_{Jeong09}-NUV_{our}) \rangle$=0.01$\pm$0.05;
NGC~4552 $\langle (FUV_{Jeong09}-FUV_{our}) \rangle$=-0.14$\pm$0.14 and 
 $\langle (NUV_{Jeong09}-NUV_{our}) \rangle$=-0.04$\pm$0.07;
NGC~5813  $\langle (FUV_{Jeong09}-FUV_{our}) \rangle$=0.11$\pm$0.08 and 
 $\langle (NUV_{Jeong09}-NUV_{our}) \rangle$=0.11$\pm$0.07;
NGC~5831  $\langle (FUV_{Jeong09}-FUV_{our}) \rangle$=-0.18$\pm$0.22 and 
$\langle (NUV_{Jeong09}-NUV_{our}) \rangle$=-0.01$\pm$0.09;
NGC~5846  $\langle (FUV_{Jeong09}-FUV_{our}) \rangle$=-0.01$\pm$0.07 and 
$\langle (NUV_{Jeong09}-NUV_{our}) \rangle$=0.02$\pm$0.04. 
 
\begin{table}
\scriptsize{
\caption{Journal of the {\it GALEX} observations}
\begin{tabular}{lllllllll}
\hline\hline
\multicolumn{1}{l}{Ident.}&
\multicolumn{1}{l}{FUV }&
\multicolumn{1}{l}{NUV } &
\multicolumn{1}{l}{Observing} \\
\multicolumn{1}{c}{} &
\multicolumn{1}{l}{Exp. Time} &
\multicolumn{1}{l}{Exp. Time} &
\multicolumn{1}{l}{program} \\
\multicolumn{1}{c}{} &
\multicolumn{1}{l}{[sec]} &
\multicolumn{1}{l}{[sec]} &
\multicolumn{1}{l}{} \\
\hline
 NGC 128  & 1499 & 1499 & GI3 089001    \\   
 NGC 777   & 1992 & 1992 & NGA\_NGC777  \\ 
 NGC 1052 & 2982 & 3834 & NGA\_NGC1052  \\ 
 NGC 1209 &1637 &2940  & GI3 087001     \\ 
 NGC 1380 & 1646 & 1646 & NGA\_NGC1380  \\ 
 NGC 1389 & 17543 & 18602 & FORNAX\_MOS08\\
 NGC 1407 & 1575 & 1575 & NGC\_NGC1407  \\ 
 NGC 1426 & 1693& 1693 & GI3 087002     \\ 
 NGC 1453 &3382 &5301  &GI3 087003      \\ 
 NGC 1521 &1662 & 11746 & GI1\_047024   \\ 
 NGC 1533 &1520 & 3152 &GI3 087004      \\ 
 NGC 1553 &2016 & 2016 & NGA\_NGC1553   \\ 
 NGC 2911 & 1605 & 1605 & GI3\_079009   \\ 	
 NGC 2962& 2297 & 2297 & MISWZN09\_24136\_0336 \\  
 NGC 2974$^{1,2}$ & 2657 & 2657 & GI1\_109006    \\ 
 NGC 3258 &2223 & 2223  &GI3 087006      \\
 NGC 3268 &2223 & 2223 &GI3 087006       \\
 NGC 3489 &1642 & 1642 &GI3 087007       \\ 
 NGC 3607 & 2468 & 2468 & GI1-079016     \\
 NGC 3818 & 1166 &1166 &GI3 087008       \\
 NGC 3962 & 2098&2098 &GI3 087009         \\
 NGC 4374$^{2}$& 1606 & 2978 & NGC\_Virgo\_MOS10 \\ 
 NGC 4552$^{2}$ & 1600 & 4770 & NGA\_Virgo\_MOS03 \\ 
 NGC 4697 & 1696 & 1696 & GI4\_085003      \\          \\
 NGC 5011 &1578&1578 &GI3 087010           \\ 
 NGC 5044 & 1696&1696 &GI3 087011          \\ 
 NGC 5363 & 16947 & 18451 & PS\_VISTA\_MOS04 \\ 
 NGC 5638 & 1704 & 1704 & MISDR1\_33739\_0535 \\ 
 NGC 5813$^{2}$ & 2539 & 2539 & GI3\_041009   \\  
 NGC 5831$^{2}$ & 2372 & 5335 & GI1\_109008   \\  
 NGC 5846$^{2}$& 1479 & 1479  &  GI3\_041010  \\  
 NGC 6868 &1700 &1700  &GI3 087012      \\  
 NGC 6958 & 3094 & 3094 & NGA\_NGC6958  \\  
 NGC 7079 &1708&1709    &GI3 087013     \\  
 NGC 7135$^{3}$ &1693 & 1693  & GI1\_059005   \\  
 NGC 7192 &1754 &1754   &GI3 087014     \\  
 NGC 7332& 1807 & 1807  & GI3\_079031   \\                                        
 IC 1459    & 1678 & 1678   & GI1\_093001\\ 					    
 IC 4296 &1701 &3365       &GI3 087015   \\  					    
 IC 5063 &2958& 2958        & GI3 087016 \\  					    
\hline												    
\end{tabular}}											    
												    
We indicate the ETGs for which a detailed UV surface photometry has been performed:		    
$^{1}$  \citet{Jeong07}; $^{2}$ \citet{Jeong09}; $^{3}$ \citet{Rampazzo07}.			    
\label{table2}											    
\end{table}											    

In addition to the UV data, we used optical     
SDSS archival data \citep{Ade08} in
the $u$ [2980-4130 \AA], g [3630-5830 \AA], $r$ [5380-7230 \AA], $i$
[6430-8630 \AA] and  $z$ [7730-11230 \AA] bands available for 14
ETGs in our sample (see Figure B.1 in Appendix  B for  SDSS composite images).
We registered the SDSS images  (corrected frames  with the `soft bias' 
of 1000 subtracted) to the corresponding {\it GALEX} NUV image 
using the IRAF tool {\tt sregister}.
We computed the SDSS magnitudes in the {\it r} band  within r$_e/8$ and  D$_{25}$\footnote{We converted 
SDSS counts to magnitudes following the recipe provided in 
{\tt http://www.sdss.org/df7/algorithms/fluxcal.html \#counts2mag}} (Table ~3). 
Figure~\ref{SDSS}, in  Appendix B, shows the surface brightness profiles 
in the $r$ band and the (FUV-$r$) and (NUV-$r$) colour profiles. 
The   integrated optical photometry will be used in Section~4.6 to
investigate the UV-optical colour magnitude relation for the
sub-sample of 14 ETGs.

\section{Results and Discussion}

\subsection{FUV and NUV morphology}

In general, the FUV emission is less extended  
than the NUV emission \citep[see e.g.][]{Gil07,Jeong07,Rampazzo07,Jeong09}. 
 
The barred lenticular galaxies NGC~1533, NGC~2962 and NGC~3489  
show rings and loops.  
NGC~2974, classified in RC3 as E4, likely
has a small scale and possibly a large scale bar. \citet{Krajnovic05}
found, with {\tt SAURON} observations, the existence of non-axisymmetric 
perturbations consistent with the presence of inner bars.

According to the different statistical estimates, rings are observed
in a significant fraction, up to 20-30\%, of lenticulars and spirals   
\cite[e.g.][]{Thilker07, Bianchi07}  and are closely
associated with non-axisymmetric structures like bars, ovals or triaxial
bulges. This fact is traditionally explained with orbits crowding and  gas
accumulation at the Lindblad resonances \citep[see e.g.][]{Buta96}. 
This kind of rings has an internally driven origin and differs both from rings 
generated by galaxy-galaxy interaction, e.g. head-on collisional rings, 
\citep[see e.g.][]{Athanassoula09} and from merging events, e.g. polar rings 
\citep{Marino09}.

In gas rich systems  the bar drives gas inflow due to its gravity torque and
could then trigger nuclear star formation and/or AGN activity 
\citep[see e.g.][]{Bournaud05}.  This does not seem the case of 
both the nuclei of NGC~1533 and of  NGC 2974 which appear
quite old with a luminosity-weighted age  $>$10 Gyr (see Table~1). 
In the case of NGC~2974, a possible explanation is suggested by the
complex gas kinematics evidenced by the work of 
\citet{Krajnovic05} which likely prevents star formation.

The NUV as well as the FUV emission in the ring structures
are not uniform and present knotty regions. 
Arm-like structures departing from the ring are 
clearly visible in NGC~2974 as well as in NGC~2962. 
\citet{Sandage79} suggest the presence of a weak spiral 
pattern in the outer lense in NGC~1533. In the SW of the NUV image 
a spiral pattern, departing from the ring, is visible. 
The NUV and FUV images show the presence of a nucleus while the bar is not
clearly visible.  The ring structure is not as regular as in 
the NIR band images (and residuals) 
shown by \citet{Laurikainen06} but is marked by bright knots
in the NE visible also in the FUV band.    

A complex loop  structure is evident in the unbarred  
lenticular Seyfert  galaxy IC~5063.  
This structure, seen nearly edge-on, seems not confined to a plane. 

Prominent dust structures are clearly visible in NGC~5363 and NGC~7192
(Figure~\ref{fig1}). 

There are seven shell galaxies in the sample, namely NGC~1553, 
NGC~2974, NGC~4552, NGC~6958, NGC~7135, NGC~7192 
and IC~1459 \citep{MC83,Tal09}. 
Only NGC~7135 shows the  peculiar shell structure visible in the optical image,
enhanced by the higher contrast provided by the UV sensitivity to the younger stellar
population  \citep{Rampazzo07}. 
In particular, the FUV image shows a faint tail, departing from the nucleus 
towards the south, which corresponds to an H$\alpha$ feature 
detected by \citet{Rampazzo05} in a Fabry-Perot study of the galaxy.

\begin{table*}
\small{
\caption{FUV, NUV {\it GALEX} and SDSS $r$ aperture photometry.}
 \begin{tabular}{lcccccccc}
\hline\hline\noalign{\smallskip}
Ident.  &   FUV$_{r_e/8}$  & NUV$_{r_e/8}$  & FUV$_{r_e/4}$ & NUV$_{r_e/4}$ &FUV$_{D25}$ & NUV$_{D25}$ & r$_{r_e/8}$ & r$_{D25}$ \\
      & [AB mag]    & [AB mag]    &   [AB mag]      & [AB mag] &  [AB mag] &[AB mag] &  [AB mag] &[AB mag]       \\
\hline      
NGC128  & 22.41$\pm$0.24  & 21.67$\pm$0.11& 20.59$\pm$0.13 & 19.82$\pm$0.07  & 18.70$\pm$0.11 & 17.70$\pm$0.06 &  &   \\
NGC777  & 19.48$\pm$0.09  & 19.21$\pm$0.05& 18.89$\pm$0.09 & 18.47$\pm$0.05  & 18.34$\pm$0.11 & 17.31$\pm$0.05 &  &   \\
NGC1052 & 19.19$\pm$0.08  & 18.48$\pm$0.04& 18.47$\pm$0.07 & 17.68$\pm$0.04  & 17.37$\pm$0.07 & 16.12$\pm$0.04 & 12.53$\pm$0.02  & 10.36$\pm$0.01 \\
NGC1209 & 21.60$\pm$0.16  & 20.86$\pm$0.07& 19.92$\pm$0.11 & 19.07$\pm$0.05  & 18.49$\pm$0.10 & 17.22$\pm$0.04 &  &   \\
NGC1380 & 20.50$\pm$0.12  & 19.72$\pm$0.05& 19.42$\pm$0.10 & 18.42$\pm$0.04  & 16.97$\pm$0.08 & 15.49$\pm$0.04 &  &   \\
NGC1389 & 22.22$\pm$0.08  & 20.85$\pm$0.04& 20.92$\pm$0.07 & 19.38$\pm$0.04  & 18.78$\pm$0.07 & 16.93$\pm$0.03 &  &   \\
NGC1407 & 18.09$\pm$0.07  & 17.73$\pm$0.04& 17.55$\pm$0.07 & 17.09$\pm$0.04  & 16.46$\pm$0.08 & 15.61$\pm$0.04 &  &   \\
NGC1426 & 21.28$\pm$0.15  & 19.92$\pm$0.06& 20.23$\pm$0.12 & 18.83$\pm$0.05  & 18.60$\pm$0.11 & 16.87$\pm$0.05 &  &   \\
NGC1453 & 20.82$\pm$0.11  & 20.34$\pm$0.06& 19.91$\pm$0.10 & 19.32$\pm$0.05  & 18.45$\pm$0.11 & 17.54$\pm$0.05 &  &   \\
NGC1521 & 21.50$\pm$0.19  & 20.14$\pm$0.07& 20.73$\pm$0.16 & 19.23$\pm$0.06  & 19.30$\pm$0.22 & 17.34$\pm$0.05 &  &   \\
NGC1533 & 19.81$\pm$0.11  & 19.18$\pm$0.05& 18.91$\pm$0.09 & 18.10$\pm$0.04  & 17.01$\pm$0.08 & 15.97$\pm$0.04 &  &   \\
NGC1553 & 18.70$\pm$0.08  & 17.39$\pm$0.04& 17.98$\pm$0.07 & 16.58$\pm$0.04  & 16.62$\pm$0.07 & 14.46$\pm$0.03 &  &   \\
NGC2911 & 22.00$\pm$0.24  & 20.53$\pm$0.09& 21.12$\pm$0.19 & 19.51$\pm$0.07  & 19.44$\pm$0.13 & 17.32$\pm$0.05 & 13.50$\pm$0.02 & 11.49$\pm$0.02 \\
NGC2962 & 21.77$\pm$0.18  & 20.83$\pm$0.08& 20.87$\pm$0.14 & 19.80$\pm$0.06  & 18.91$\pm$0.10 & 17.81$\pm$0.05 & 14.03$\pm$0.02 & 11.78$\pm$ 0.02\\
NGC2974 & 20.25$\pm$0.10  & 19.39$\pm$0.05& 19.48$\pm$0.08 & 18.51$\pm$0.05  & 17.65$\pm$0.08 & 16.34$\pm$0.04 &  &   \\
NGC3258 & 19.68$\pm$0.10  & 19.48$\pm$0.06& 18.91$\pm$0.09 & 18.10$\pm$0.04  & 18.19$\pm$0.14 & 17.07$\pm$0.06 &  &   \\
NGC3268 & 20.28$\pm$0.12  & 19.69$\pm$0.06& 19.73$\pm$0.11 & 18.99$\pm$0.06  & 18.43$\pm$0.17 & 16.86$\pm$0.05 &  &   \\
NGC3489 & 19.29$\pm$0.09  & 17.88$\pm$0.04& 18.68$\pm$0.08 & 17.21$\pm$0.04  & 16.86$\pm$0.07 & 15.13$\pm$0.04 & 12.41$\pm$0.02 &  10.14$\pm$0.01 \\
NGC3607 & 18.66$\pm$0.07  & 17.89$\pm$0.04& 17.61$\pm$0.06 & 16.84$\pm$0.04  & 16.65$\pm$0.07 & 15.38$\pm$0.04 & 12.07$\pm$0.02 &  9.94$\pm$0.01\\
NGC3818 & 21.07$\pm$0.15  & 20.11$\pm$0.08& 20.23$\pm$0.12 & 19.10$\pm$0.06  & 19.16$\pm$0.14 & 17.41$\pm$0.05 &  &  \\
NGC3962 & 20.11$\pm$0.10  & 19.20$\pm$0.05& 19.32$\pm$0.09 & 18.25$\pm$0.05  & 17.89$\pm$0.09 & 16.34$\pm$0.04 &  &   \\
NGC4374 & 18.35$\pm$0.08  & 17.59$\pm$0.04& 17.61$\pm$0.07 & 16.76$\pm$0.04  & 16.32$\pm$0.07 & 15.13$\pm$0.03 & 11.48$\pm$0.01 & 9.52$\pm$0.01 \\
NGC4552 & 17.95$\pm$0.07  & 17.86$\pm$0.04& 17.24$\pm$0.07 & 16.94$\pm$0.03  & 16.17$\pm$0.07 & 15.20$\pm$0.03 & 12.29$\pm$0.02 & 9.39$\pm$0.01\\
NGC4697 & 18.73$\pm$0.08  & 17.43$\pm$0.04& 18.04$\pm$0.08 & 16.61$\pm$0.04  & 16.58$\pm$0.08 & 14.83$\pm$0.04 & 11.34$\pm$0.01 & 9.35$\pm$0.01\\ 
NGC5011 & 21.69$\pm$0.21  & 20.41$\pm$0.08& 20.74$\pm$0.17 & 19.33$\pm$0.07  & 22.25$\pm$0.54 & 17.19$\pm$0.06 &  &   \\
NGC5044 & 18.91$\pm$0.09  & 18.41$\pm$0.05& 18.23$\pm$0.09 & 17.61$\pm$0.05  & 17.39$\pm$0.10 & 16.39$\pm$0.05 &  &   \\
NGC5363 & 18.83$\pm$0.06  & 18.28$\pm$0.03& 18.08$\pm$0.06 & 17.32$\pm$0.03  & 17.03$\pm$0.06 & 15.71$\pm$0.03 & 12.60$\pm$0.02 & 9.87$\pm$0.01 \\
NGC5638 & 20.93$\pm$0.14  & 19.86$\pm$0.06& 20.04$\pm$0.11 & 18.68$\pm$0.05  & 18.49$\pm$0.10 & 16.79$\pm$0.04 & 13.57$\pm$0.02 & 11.19$\pm$0.02\\
NGC5813 & 19.78$\pm$0.09  & 18.77$\pm$0.05& 19.11$\pm$0.08 & 17.99$\pm$0.04  & 18.22$\pm$0.10 & 16.84$\pm$0.04 & 14.95$\pm$0.04 & 10.85$\pm$0.01\\
NGC5831 & 21.11$\pm$0.13  & 20.09$\pm$0.05& 20.29$\pm$0.11 & 19.11$\pm$0.04  & 18.95$\pm$0.13 & 17.16$\pm$0.04 & 13.60$\pm$0.02 & 11.22$\pm$0.02\\
NGC5846 & 18.88$\pm$0.09  & 18.28$\pm$0.05& 18.11$\pm$0.08 & 17.43$\pm$0.04  & 17.12$\pm$0.09 & 15.95$\pm$0.04 & 12.29$\pm$0.02 & 10.07$\pm$0.02 \\
NGC6868 & 20.16$\pm$0.11  & 19.26$\pm$0.06& 19.26$\pm$0.09 & 18.31$\pm$0.05  & 18.05$\pm$0.12 & 16.41$\pm$0.05 &  &   \\
NGC6958 & 21.24$\pm$0.13  & 19.95$\pm$0.06& 20.31$\pm$0.10 & 18.86$\pm$0.05  & 18.72$\pm$0.11 & 16.84$\pm$0.04 &  &   \\
NGC7079 & 21.84$\pm$0.18  & 20.02$\pm$0.07& 21.10$\pm$0.14 & 19.24$\pm$0.06  & 19.49$\pm$0.15 & 16.97$\pm$0.04 &  &   \\
NGC7135 & 19.73$\pm$0.10  & 19.52$\pm$0.06& 19.31$\pm$0.09 & 18.88$\pm$0.05  & 18.10$\pm$0.10 & 17.07$\pm$0.05 &  &   \\
NGC7192 & 21.09$\pm$0.15  & 19.87$\pm$0.06& 20.19$\pm$0.12 & 18.85$\pm$0.05  & 18.53$\pm$0.11 & 16.98$\pm$0.04 &  &   \\ 
NGC7332 & 22.20$\pm$0.20  & 20.39$\pm$0.07& 20.92$\pm$0.14 & 19.02$\pm$0.05  & 18.69$\pm$0.12 & 16.56$\pm$0.04 & 13.50$\pm$0.02 & 10.87$\pm$0.01\\
IC1459  & 18.39$\pm$0.07  & 18.04$\pm$0.04& 17.75$\pm$0.07 & 17.25$\pm$0.04  & 16.49$\pm$0.07 & 15.45$\pm$0.04 &  &   \\
IC4296  & 19.56$\pm$0.10  & 18.84$\pm$0.06& 18.82$\pm$0.09 & 18.00$\pm$0.05  & 17.68$\pm$0.12 & 16.34$\pm$0.05 &  &   \\
IC5063  & 20.60$\pm$0.10  & 19.61$\pm$0.05& 19.79$\pm$0.09 & 18.72$\pm$0.05  & 18.17$\pm$0.09 & 17.05$\pm$0.04 &  &   \\
\hline
\end{tabular}

{  Note: Magnitudes are not corrected for galactic extinction.}
} 
\label{tab3}
\end{table*}

\begin{figure}
\begin{tabular}{c}
\vspace{-0.8cm}
\psfig{figure=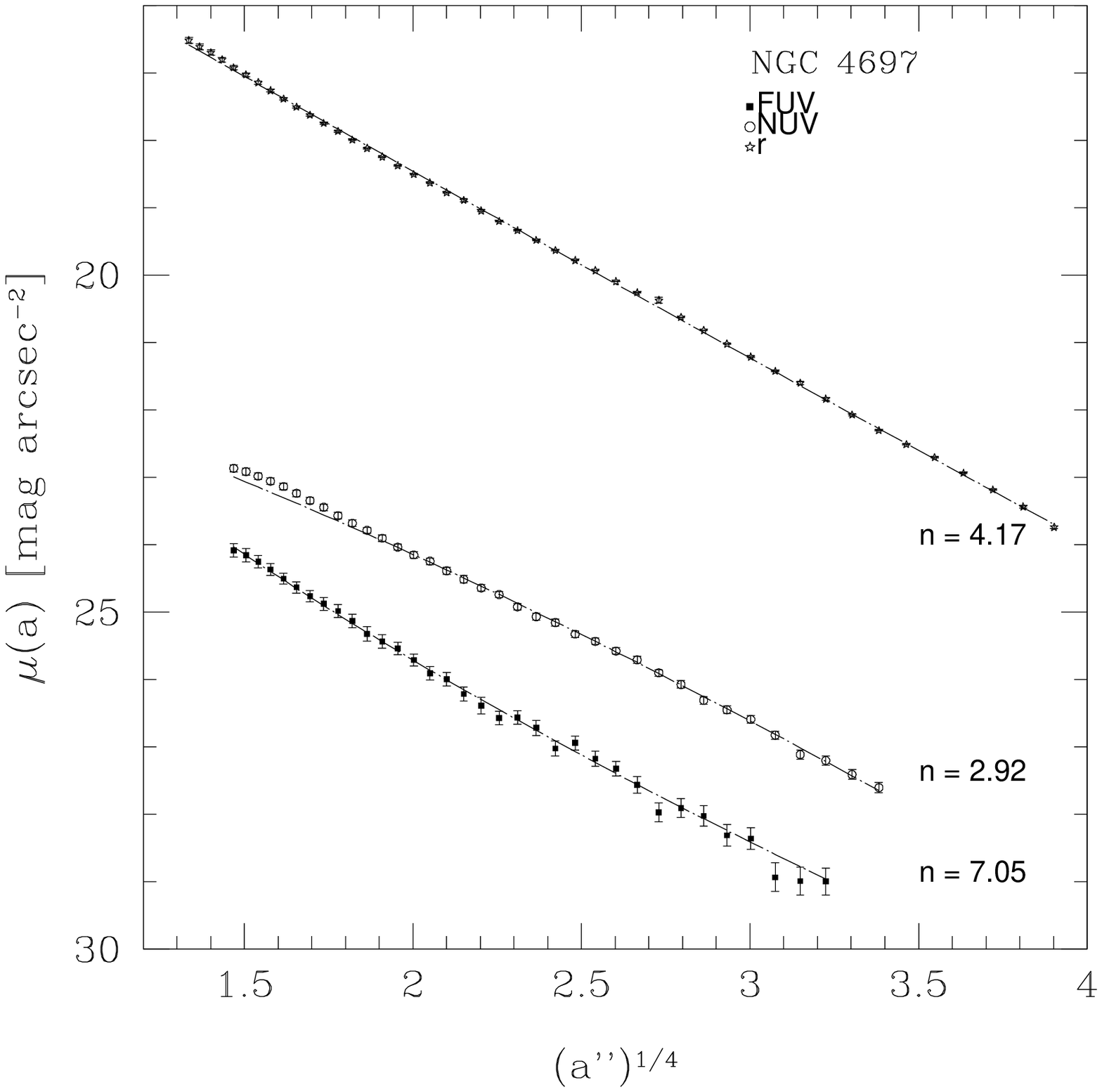,width=7.5cm} \\
\vspace{-0.6cm}
\psfig{figure=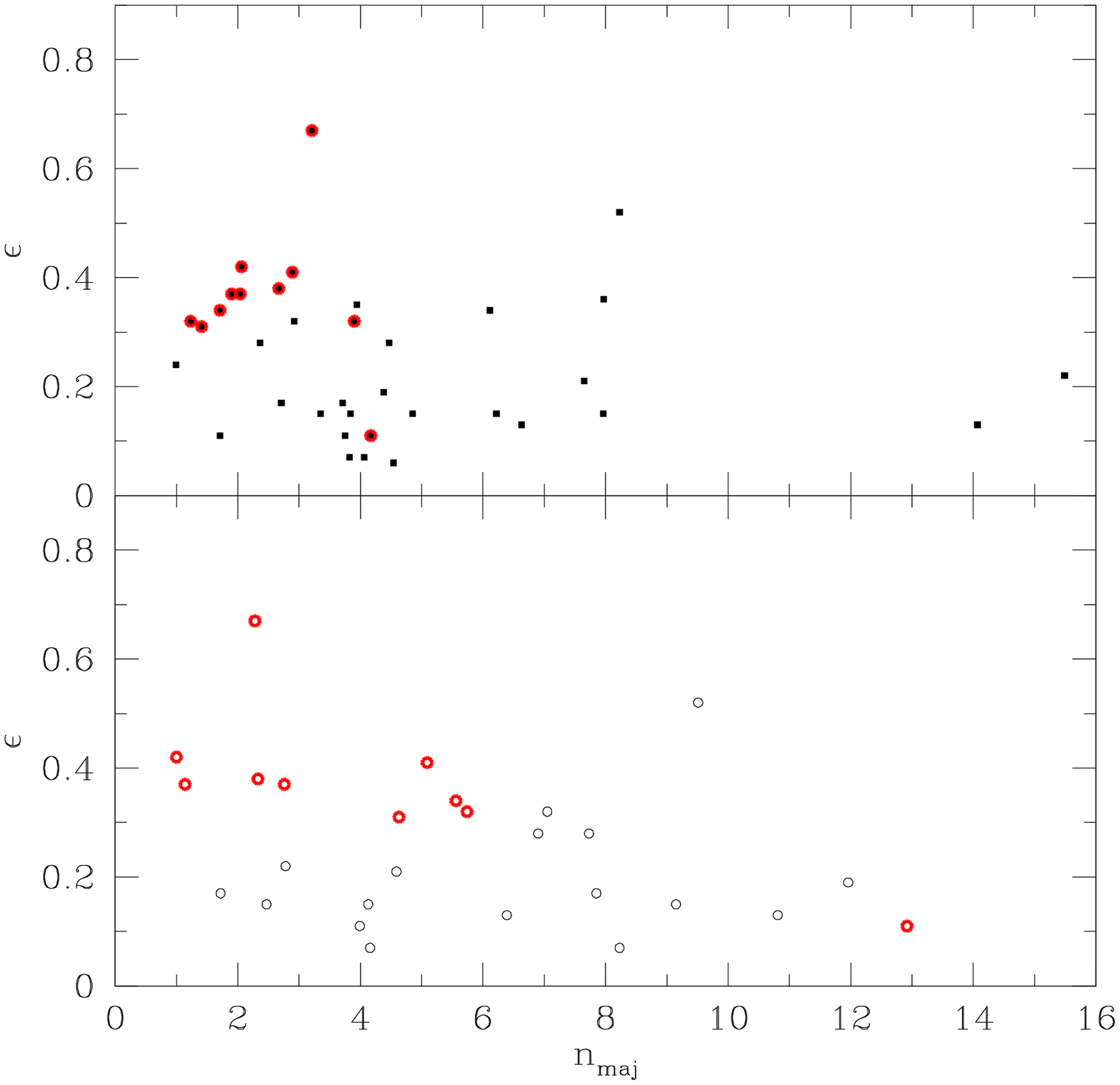,width=6.65cm}   \\
 \psfig{figure=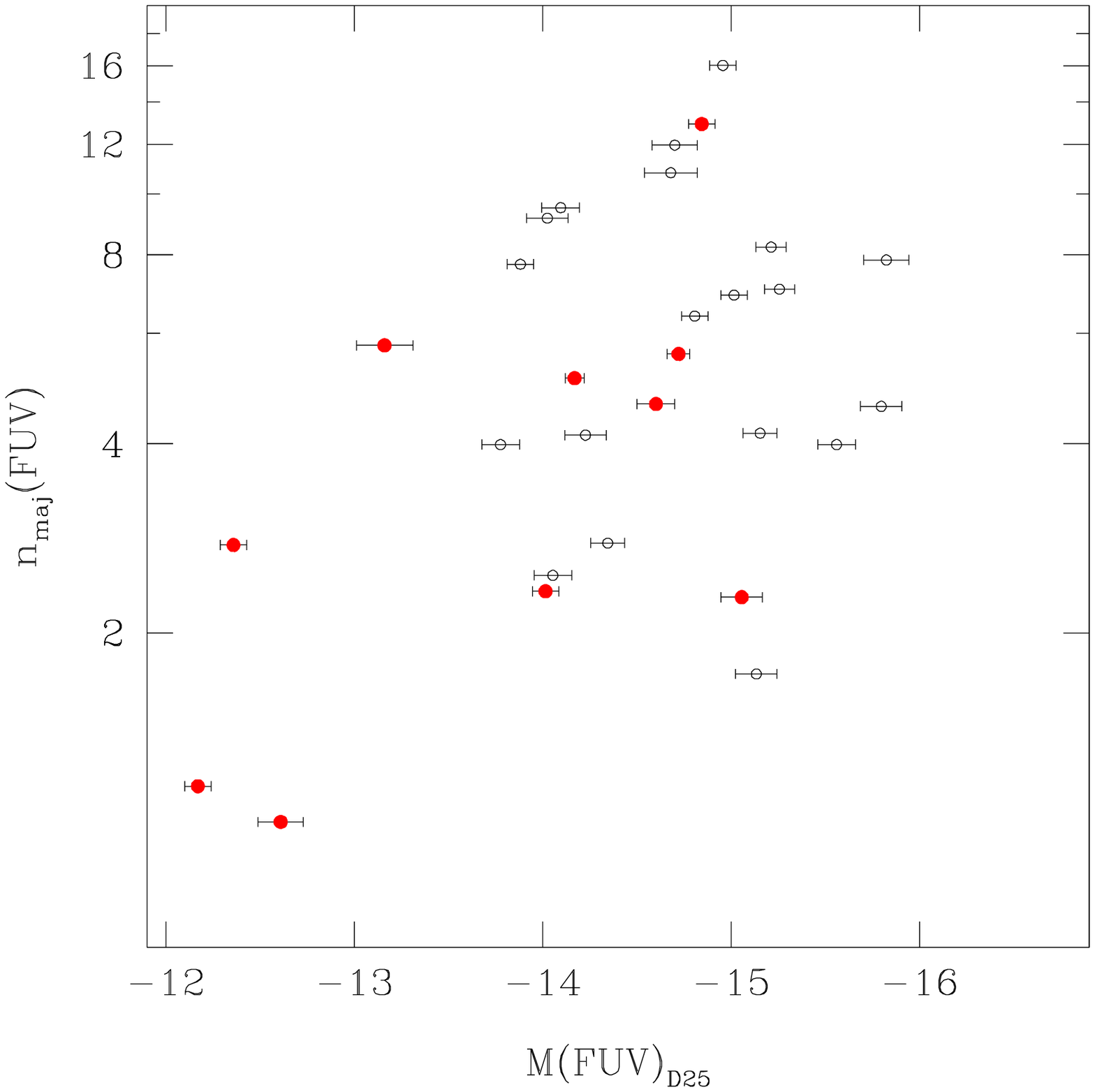,width=7.5cm}\\
 \end{tabular}
 \vspace{-0.5cm}
\caption{Top panel: example of fitting with a Sersic law  of the
observed optical and {\it GALEX} surface brightness profiles of NGC~4697 along the major axis. 
The index of the Sersic law, $n$, derived from the best fitting of the  $r$ (stars), 
of the NUV (open circles) and  FUV (filled squares) surface brightness profiles is indicated. 
Middle panel: plot of the average ellipticity $\epsilon$ vs. $n$ indices calculated 
in the NUV (top) and FUV  (bottom) bands.  Large red circles mark the S0 galaxies.  
Bottom panel:  the FUV absolute magnitude integrated 
within the optical $D_{25}$  diameter is plotted vs. $n$.   
Data are provided in Table~3 and Table~\ref{table5}
} 
 \label{fig3}
\end{figure}

\subsection{The UV surface brightness profiles}

To quantitatively describe the shape of the UV 
surface brightness profiles we use the Sersic law \citep{Sersic68}.
The Sersic profile is a generalization of the de Vaucouleur
law with $\mu(r) \sim r^{1/n}$,
where $n$ is a free parameter,  named the
Sersic index. The profile is thus sensitive to structural
differences between  ETGs and 
provides a better fitting to real galaxy profiles. 

The Sersic  $r^{(1/n)}$ description is suited to represent 
the family of ETGs surface brightness profiles:
for $n=1$ the formula describes a simple exponential profile,
while for $n=4$ it expresses the de Vaucouleurs law.    

For profile fitting we use  the formalism and the methods
described in \citet{Caon93}. We adopt:   
$$ \mu(r) = A + B~r^{1/n}$$  which is connected
to  the logarithmic form of the Sersic law 
$$ \mu(r) = \mu_e + c_n [(\frac{r}{r_e})^{1/n} - 1] $$
through the effective parameters $\mu_e=A+c_n$ and $r_e=(c_n/B)^n$.  
The estimate of both the above parameters is then model dependent.

In Table~\ref{table5} we provide for each galaxy the value of $n$,
the coefficient $A$ and $B$ of the interpolation, and the surface brightness
interval $\mu_{start}$  (to avoid the
influence of the PSF in the centre) and $\mu_{end}$ corresponding to the truncation of the profile when the 
error exceeds 0.3 mag~arcsec$^{-2}$. An inner
cut-off of 3 times the PSF provides a safe determination of $n$ \citep[see
also][]{Caon93, Brown03}.
     
The $n$ index with the smallest $\sigma_{bf}$  
scatter is selected and reported in Table~\ref{table5}.  
\citet{Caon93}  describe the uncertainty associated 
with the measure of $n$.  They compute the variation of  $n$, $\Delta n$,
if the rms scatter of the (O-C) residual exceeds by 25\% the value of the best fitting.
For $n\leq10$, $\Delta n \simeq 0.25 n$ while for larger $n$ the uncertainty
may easily exceed 2, i.e. $n\pm2$. 

 In Table~\ref{table5} we describe the quality of the whole 
fitting  as judged by visual inspection.  
The profile quality is classified ``poor" in  the presence of distinct, intrinsic systematic deviations 
of the luminosity profile from the  best fit Sersic law. 
When  the fitting interval 
is too small (less than 3 mag/arcsec$^{-2}$), the result is not reported in 
Table~\ref{table5}.
We do not fit ETGs whose profiles are dominated by the presence of a ring,
namely NGC~1533, NGC~2962, NGC~2974, IC~5063. In the case of 
NGC~3489 the fitting is quite poor, although fitting parameters are provided.
The FUV surface brightness profiles are generally more noisy
than those in the NUV band and  the fitting  results have higher uncertainties. 
 
In Figure~\ref{fig3} (top panel) we plot an example of the Sersic fit  
in the FUV, NUV and $r$ bands. We notice that the derived  index
$n$ is significantly different in the three bands for some 
ETGs.   
Some large values of $n$ are found
also in optical bands where surface brightness profiles are best fitted
over a much larger magnitude range from the nucleus to the galaxy outskirts.  
For example, our  largest Sersic index  $n(FUV)$=16.02 of the shell galaxy NGC 4552,
for which however we find $n$ = 4.5 and 6 in NUV and {\it r}, 
is comparable to the \citet{Caon93}  value of $n_{maj}$=13.87 in the B band. For the same galaxy 
\citet{Kormendy09} derive in the V band $n=9.22^{+1.13}_{-0.83}$, which 
differs by $>$ 3$\sigma$  from  our {\it r} value. On the other hand, for NGC~4374, another
slow rotator like NGC~4552 but without perturbation signatures (see Paper~II 
on-line notes), we obtain $n\approx6.5$ in all bands,  consistent
within the errors  with $n=7.9^{+0.71}_{-0.56}$ obtained by \citet{Kormendy09} 
in the V-band.

The values of $n$ in the  UV  bands range from 1 to 16
just as in the  optical \citep[e.g.][]{Caon93} and in the NIR 
\citep{Brown03} samples of ETGs.  In our SDSS sample
the range of $n$ is smaller,  from 2.55 to 9.41, 
comparable to the range of $n$ = 1.40--11.84 obtained in the V band 
by \citet{Kormendy09} for a larger sample of Virgo galaxies. 
The median of the $n$ distribution is 3.83, 5.09, 3.52 for NUV, FUV and $r$ bands,
while averages ($\pm 1\sigma$ errors) are 4.52$\pm$3.16,
5.96$\pm$3.66, 4.27$\pm$1.95, respectively. 

The average galaxy ellipticity $\epsilon$, does not correlate with
the Sersisc index (see Figure~\ref{fig3}
middle panel), although more flattened galaxies tend to have 
lower values of $n$, as noticed by \citet{Caon93}.   
Lenticular galaxies, i.e. truly disk galaxies, tend to have
$n$ values typically lower than 5 in our sample.

In the bottom panel of Figure~\ref{fig3} we plot the absolute FUV
magnitudes within  the D$_{25}$ aperture versus the $n$ value
from the FUV profile. The plot shows a  weak correlation (correlation 
coefficient 0.350) and a rather large dispersion. The sense of the 
correlation is that $n$ becomes larger with increasing total luminosity
as observed in several samples. \citet{Brown03} (their Figure~4), 
\citet{Coenda05}  (their Figure~9) found a  
large scatter when the $n$ values are plotted vs. 
the total absolute K, R and V band magnitudes. 
A correlation, although weak, between $n$ 
and the bulge K total absolute magnitude was found by   
\citet{Andredakis95}. \citet{Caon93} and 
\citet{Prugniel97} found that $n$ and the absolute B, 
$M_B$, band magnitudes are significantly correlated. 
\citet{Prugniel97} suggest that the non-homology,
mapped by the Sersic index, may contribute to the tilt of the Fundamental Plane. 
They show that   $n$ correlates with the residual of the ETGs Fundamental Plane, 
i.e. that the non-homology  in the ETGs structure has a measurable 
effect on their scaling relations. 

\citet{Kormendy09} parametrized the surface brightness profiles
of Virgo Es with the Sersic law.  They found that the  dichotomy 
between core Es (generally slowly rotating, with a relatively high anisotropic 
velocity distribution and boxy isophotes) and `light Es' (fast rotators, with
disky isophotes) were also reflected in their Sersic index. Core Es tend to have 
$n > 4$ while `light Es' have $n \simeq 3\pm1$. \citet{Kormendy09} 
suggest that the two families of Es have a different origin since core Es tends to be
$\alpha$-enhanced relative to `light Es'.     

\begin{figure}
\psfig{figure=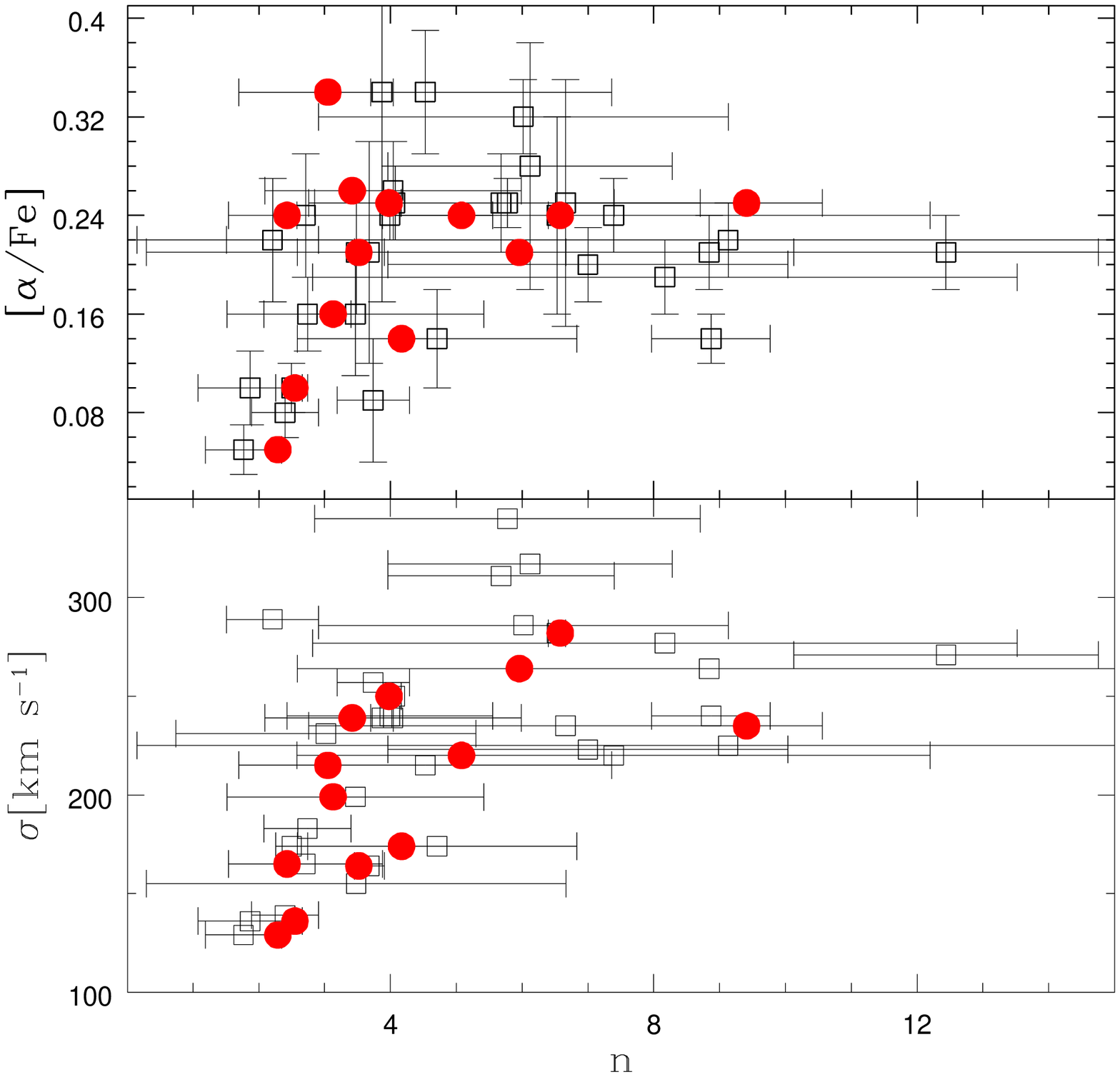,width=8.5cm} 
\caption{({\it Top panel}) The Sersic index $n$ vs. nuclear [$\alpha$/Fe]. 
 Open squares indicate the values obtained
 averaging  the Sersic indices derived in FUV, NUV and r. 
 Filled dots are for the n values derived from the r-band alone.  
 ({\it Bottom panel}) The average Sersic index $n$ vs. the galaxy central velocity 
dispersion (Table~1).  
 } 
\label{fig4}
\end{figure}

In the top panel of Figure~\ref{fig4} we plot the average $n$ 
index of a galaxy (the value and the
error are obtained averaging over the UV and $r$ bands) 
versus the corresponding measure of the $\alpha$-enhancements 
within r$_e$/8 (column 10 in Table 1).  
ETGs with $n$ $>$ 4 tend to have [$\alpha$/Fe] values
larger than 0.15, while those with $n$ $<$ 4 span
a wider range of [$\alpha$/Fe], from $\sim$ 0.04 to  $\sim$ 0.34.
This trend is confirmed if we consider the Sersic indices obtained
from the r-band photometry. Galaxies with n=3 $\pm$ 1 are mostly
fast rotators, S0 galaxies. Among them there are the galaxies with the
younger luminosity-weighted ages.

 In the bottom panel of Figure~\ref{fig4} we plot the Sersic index
versus the galaxy central velocity dispersion reported in Table~1.
As reported in Section~2, the ETGs velocity dispersion is 
a proxy of the galaxy mass \citep{Cappellari06}. The tight 
relation between the galaxy velocity dispersion and [$\alpha$/Fe]  is 
well described in \citet[][and reference therein]{Clemens06,Clemens09},
where it is shown that it is also independent from the galaxy environment,
as well as the velocity dispersion vs. metallicity relation. \citet{Clemens09}
argue that the timing process of formation of ETGs is determined by
the environment, while the details of the process of star formation,
which has built up the stellar mass, are entirely regulated by the halo mass.
 We suggest that the relation between the Sersic index, a global indicator of the 
galaxy structure, and  [$\alpha$/Fe] shown in Figure~\ref{fig4} likely reflects 
the more basic relation between this latter variable and the galaxy mass. 

\begin{table*}
\caption{Result of the fit in the FUV and NUV {\it GALEX} and SDSS $r$ bands } 
\begin{tabular}{lccccccc} 
	\hline \hline
Ident.      & $n_{maj} FUV$& $\sigma_{bf}$ & $\mu_{start} $& $\mu_{end}$&  A   &   B  & Quality  \\
            & $n_{maj} NUV$ &              &             &              &       &       &          \\
            &  $n_{maj} r$   &               &             &              &       &       &          \\
              &             &              & [mag/arcsec$^{-2}$]      & [mag/arcsec$^{-2}$]      &          \\
\hline
 NGC~128          &  2.28           & 0.076        & 25.70      &  28.70       &  22.37 & 1.344 & G      \\
   & 3.21            & 0.044       & 25.03       & 28.50        & 19.59 & 23.29  & G      \\
NGC~777           & 4.59           &  0.118        & 26.60     & 28.70        &  16.52 & 5.775  & P     \\
   & 7.65           & 0.126        & 26.05      &  28.82       & 12.54  & 9.445  & P     \\
NGC~1052           & 7.73           & 0.073        & 25.70       & 29.2         & 12.26  & 9.760  & F \\
 & 2.36           & 0.060          & 24.60     & 28.50        &  21.91 & 0.941  & F     \\
           & 3.05          & 0.034       & 17.40       & 23.80        & 14.14  & 2.090   & G  \\
NGC~1209           & 9.51           & 0.097        & 26.10      & 28.70        & 8.73   & 13.411  & F \\
 & 8.23           & 0.050        & 25.00      & 28.50        & 11.47  & 9.893   & G \\
NGC~1380           & 5.09           & 0.063        & 25.20      & 28.20        & 17.81  & 4.603   & P \\
  & 2.89          & 0.071        & 24.20      & 29.10         & 20.50 & 1.441   & G \\
NGC~1389          & 2.76            & 0.048         & 26.60     & 29.50        & 21.54  & 2.126  & G \\
 & 2.04           & 0.090         & 25.30     & 29.30        & 21.89  & 0.897   & G \\
NGC~1407          & 8.23           & 0.092        & 25.00     & 28.00         & 11.74   & 9.820  & P \\
	   & 3.82          & 0.043         & 24.70     & 27.80        & 19.81  & 2.429   & F \\
NGC~1426            &    --          &       --    &     --      &    --           &   --     &   --      & \\
  & 6.11          & 0.049        & 25.20     & 28.00         & 15.11  & 6.627   & F\\
NGC~1453            & 1.72        & 0.104        & 26.20      & 28.40         & 23.37   & 0.675 & P \\                     
  & 2.71          & 0.070        & 25.60     & 28.00         & 22.06  & 1.345   & P\\
NGC~1521            &    --          &       --    &     --      &    --           &   --     &   --      & \\
  & 2.97         & 0.054         & 25.60     & 28.00         & 21.09  & 1.836  & G \\
NGC~1533 &    --          &       --    &     --      &    --           &   --     &   --      & \\
           &    --          &       --    &     --      &    --           &   --     &   --      & \\
 NGC~1553          & 2.33          & 0.071       &  25.20      & 28.70          & 22.75    & 0.892    & F \\
 & 2.67           &  0.092      & 24.10      & 28.10           & 21.12   & 1.052    & G \\	   
NGC~2911           &    --          &       --    &     --      &    --           &   --     &   --      & \\
  & 3.90          & 0.049      & 26.24       & 27.50           & 21.93   & 2.128    & G \\
           & 9.41          & 0.050       &  18.54      & 24.50          & 5.94     & 10.893  & F \\
NGC~2962 &    --          &       --    &     --      &    --           &   --     &   --      & \\
           &    --          &       --    &     --      &    --           &   --     &   --      & \\
	     &    --          &       --    &     --      &    --           &   --     &   --      & \\
NGC~2974 &    --          &       --    &     --      &    --           &   --     &   --      & \\
           &    --          &       --    &     --      &    --           &   --     &   --      & \\
	     &    --          &       --    &     --      &    --           &   --     &   --      & \\
 NGC~3258          & 10.81         &  0.063      &  25.90     & 27.80          & 3.99     & 17.643   & P \\
  & 14.07         & 0.070      &   25.69     & 28.11          & 1.46     & 19.971   & P\\
NGC~3268       &    --          &       --    &     --      &    --           &   --     &   --      & \\
  & 0.99         & 0.036        & 25.45      & 26.9           & 24.52    & 0.066    & P\\
 NGC~3489          & 1.14       & 0.196        & 25.28     & 29.00           & 23.82    & 0.130    & P \\
  & 1.90       & 0.087       & 23.71     & 28.00           & 20.95    & 0.632    & P \\
           & 2.29       & 0.136        & 16.70     & 24.00           & 14.51    & 1.272    & F\\
NGC~3607           & 12.92      & 0.067        & 25.50      & 27.60           &-1.47      & 21.61  & P\\
  & 4.17       & 0.015       & 24.90      & 28.00            & 18.48    & 2.990   & G \\
           & 5.08       & 0.035        & 17.18      & 23.30           & 10.78     & 4.879  & G \\
NGC~3818           &    --          &       --    &     --      &    --           &   --     &   --      & \\
  & 7.97      & 0.098        & 25.70      & 28.50           & 14.13     & 8.282  & F \\
NGC~3962           & 2.78       & 0.115       & 26.20      & 28.30            & 21.51     & 1.898   & P\\
  & 15.49        & 0.054     & 25.50     & 27.50            &  2.36      & 19.167 & P\\
NGC~4374            & 6.39      & 0.072      & 25.00       & 28.50            & 14.84     & 6.844  & G \\
  & 6.63      & 0.030       & 24.00      & 28.00            & 14.23     & 6.569   & G\\
            & 6.58      & 0.029      & 16.60       & 23.00            & 8.19      & 6.801  & G\\
 NGC~4552          &  16.02    & 0.088      & 25.00       & 28.90             & -6.24    & 26.813   & F\\
  & 4.54      & 0.034      & 24.40       & 28.00            & 17.42     & 3.923  & G\\
           &  5.96     &  0.043      & 17.00      & 25.00             & 8.49     & 6.697   & G \\
NGC~4697           & 7.05       & 0.083      & 25.40       & 28.40            & 15.28     & 7.041  & G\\
 & 2.92       & 0.032      & 24.00       & 27.20             & 20.82    & 1.287  & G \\
           & 4.17       & 0.033      & 16.80       & 23.80            & 12.63     & 2.999  & G\\
NGC~5011           &    --          &       --    &     --      &    --           &   --     &   --      & \\
 & 7.96       & 0.082       & 25.45      & 27.00            & 13.27     & 8.72   & F\\
NGC~5044         & 3.99      & 0.100      & 25.20        & 27.80           & 18.39     & 3.806   & F\\
  & 3.75      & 0.037      & 25.20        & 27.3            & 19.98     & 2.500   & G \\
\hline 
\end{tabular}
\label{table5}
 \end{table*}

\begin{table*}
	\addtocounter{table}{-1}
\caption{Continued} 
\begin{tabular}{lccccccc} 
	\hline \hline
Ident.      & $n_{maj} FUV$& $\sigma_{bf}$ & $\mu_{start} $& $\mu_{end}$&  A   &   B  & Quality  \\
            & $n_{maj} NUV$ &              &             &              &       &       &          \\
            &  $n_{maj} r$   &               &             &              &       &       &          \\
              &             &              &  [mag/arcsec$^{-2}$]     & [mag/arcsec$^{-2}$]      &          \\
\hline
NGC~5363            & 5.56      & 0.107       & 25.20         & 30.00          & 16.42     & 5.789   & P\\
  & 1.71      & 0.086      & 24.41        & 29.10           & 23.18     & 0.3250  & P\\
           & 3.13      & 0.080       & 17.21        & 24.5            & 14.04     & 2.010   & P \\
NGC~5638           & 3.99      & 0.180       &  26.95       & 29.40           & 21.37     & 2.927   & P\\
 & 1.71      & 0.078       & 25.20        & 28.40           & 23.48     & 0.395   & P\\
           & 2.43      & 0.077       & 18.17        & 24.80           & 15.76     & 1.405   & G\\
NGC~5813           & 2.47      & 0.115      & 26.00        & 28.50            & 22.81     & 1.203   & P\\
 & 6.22      & 0.048       & 25.20        & 28.30           & 17.190    & 5.158   & F \\
           & 3.42      & 0.100      & 17.74        & 24.4             & 14.39     & 2.325   & P \\
NGC~5831           &    --          &       --    &     --      &    --           &   --     &   --      & \\
 & 3.84      & 0.068     & 25.50          & 28.70           & 19.65     & 2.913   & G\\
           & 3.52      & 0.047     & 18.30    & 25.00      & 14.11 & 28.76 & G\\
NGC~5846           & 4.16      & 0.128     & 25.40    & 28.80        & 17.64   & 4.179  & P\\
 & 4.06      & 0.056     & 24.81    & 27.80        & 18.88   & 3.052  & G\\
           & 3.98      & 0.027     & 17.90    & 23.40        & 13.76   & 2.882  & G\\
NGC~6868           & 11.96    & 0.079      & 26.00    & 28.00        & 4.04     & 17.942 & P\\
 & 4.38      & 0.035     & 25.20    & 27.80        & 18.86    & 3.349  & G\\
NGC~6958           &  9.15    & 0.113      & 27.10    & 29.20        & 9.51     & 13.310 & P\\
  & 4.85     & 0.084      & 25.40   &28.00         & 18.06    & 4.203   & F\\
NGC~7079           & 5.74    & 0.141       & 27.20    & 28.80        & 17.36    & 6.471  & P\\
  & 1.23    & 0.081       & 25.20   & 29.20        & 24.23    & 0.137  & P\\
NGC~7135           & 4.63    & 0.052      & 26.50     & 28.30        & 19.82    & 3.995  & P\\
 & 1.41    & 0.099       & 25.80    & 28.80        & 24.69    & 0.183  & P\\
NGC~7192           & 4.13    & 0.150      & 27.10     & 28.90        & 21.28    & 3.229   & P\\
  & 3.35   & 0.087      & 25.60     & 28.80        & 20.38    & 2.367   & F\\
NGC~7332    & 1.00    & 0.086   & 25.60       & 28.60        & 25.02    & 0.066   & F\\
   & 2.06    & 0.046     & 23.70     & 28.20        & 20.93    & 0.758   & F\\	      
            & 2.55    & 0.082   & 16.67       & 24.40        & 13.81    & 1.580   & F\\
IC~1459           &  6.90            &  0.057     & 25.30       & 28.50        & 14.03 & 7.894  & G       \\
   &  4.47            &  0.033     & 24.65       &  28.00       & 18.48 & 3.410  & G       \\
IC~4296          &  7.85            & 0.101       & 26.50       & 28.43        & 12.17  & 10.203  & P      \\
     &  3.71            &  0.052     & 25.30       & 27.51        & 20.31 & 2.432  & F       \\
IC~5063  &    --          &       --    &     --      &    --           &   --     &   --      & \\
          &    --          &       --    &     --      &    --           &   --     &   --      & \\
\hline 
\end{tabular}
  \end{table*}

\begin{figure*}
\begin{tabular}{cc}  
\psfig{figure=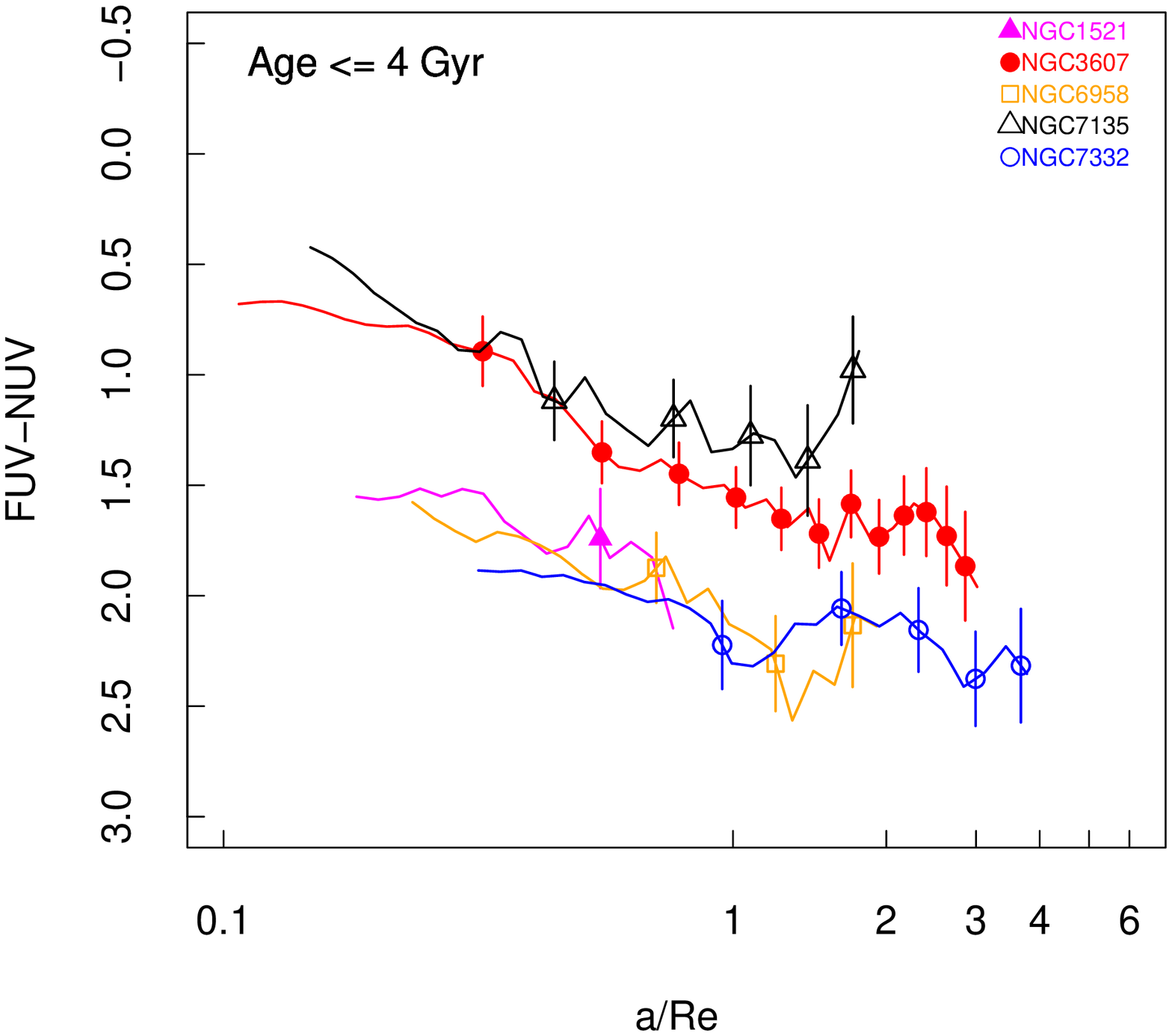,width=8.5cm} 
\vspace{-0.8cm}
\psfig{figure=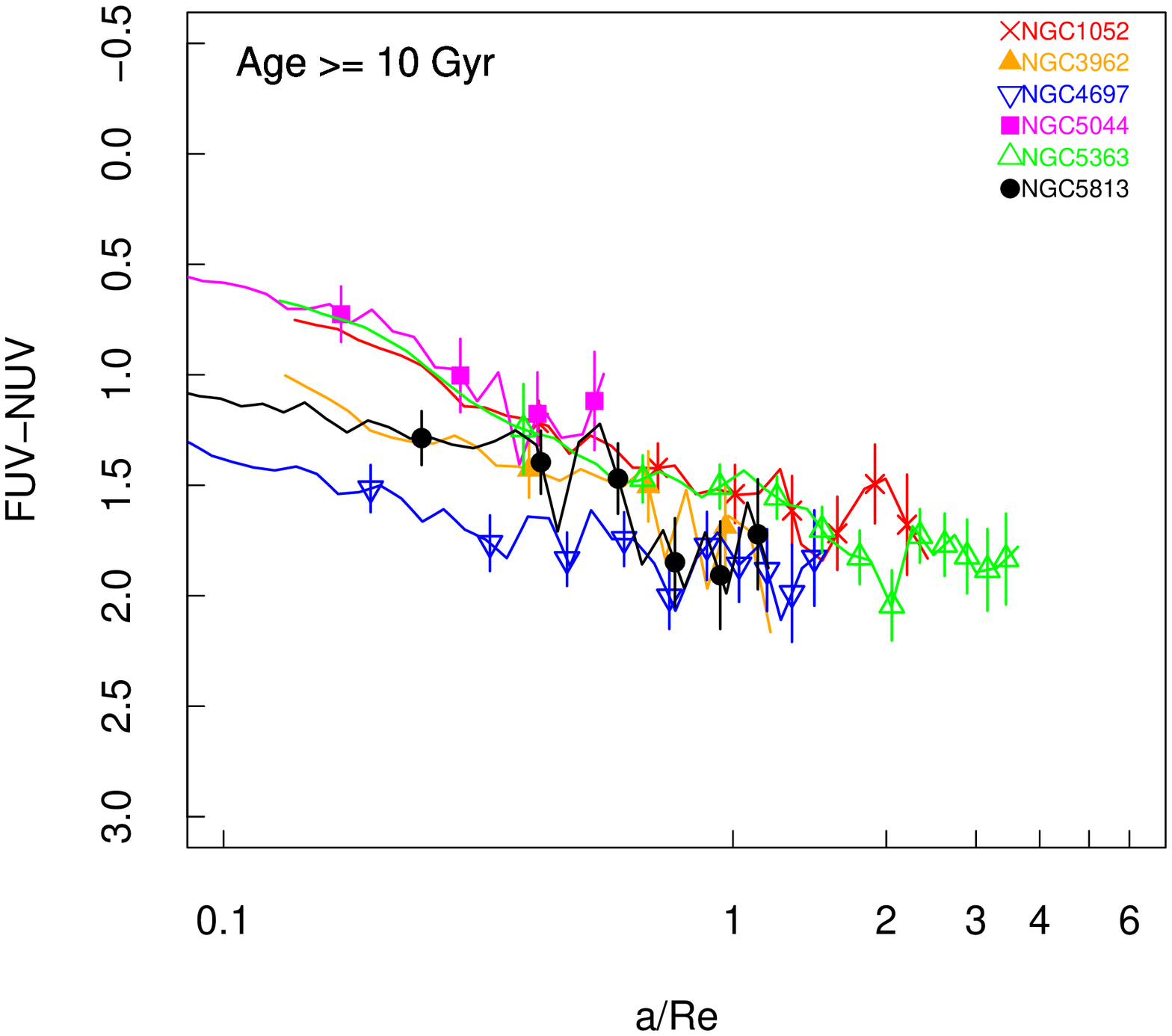,width=8.5cm} \\
\hfill
\psfig{figure=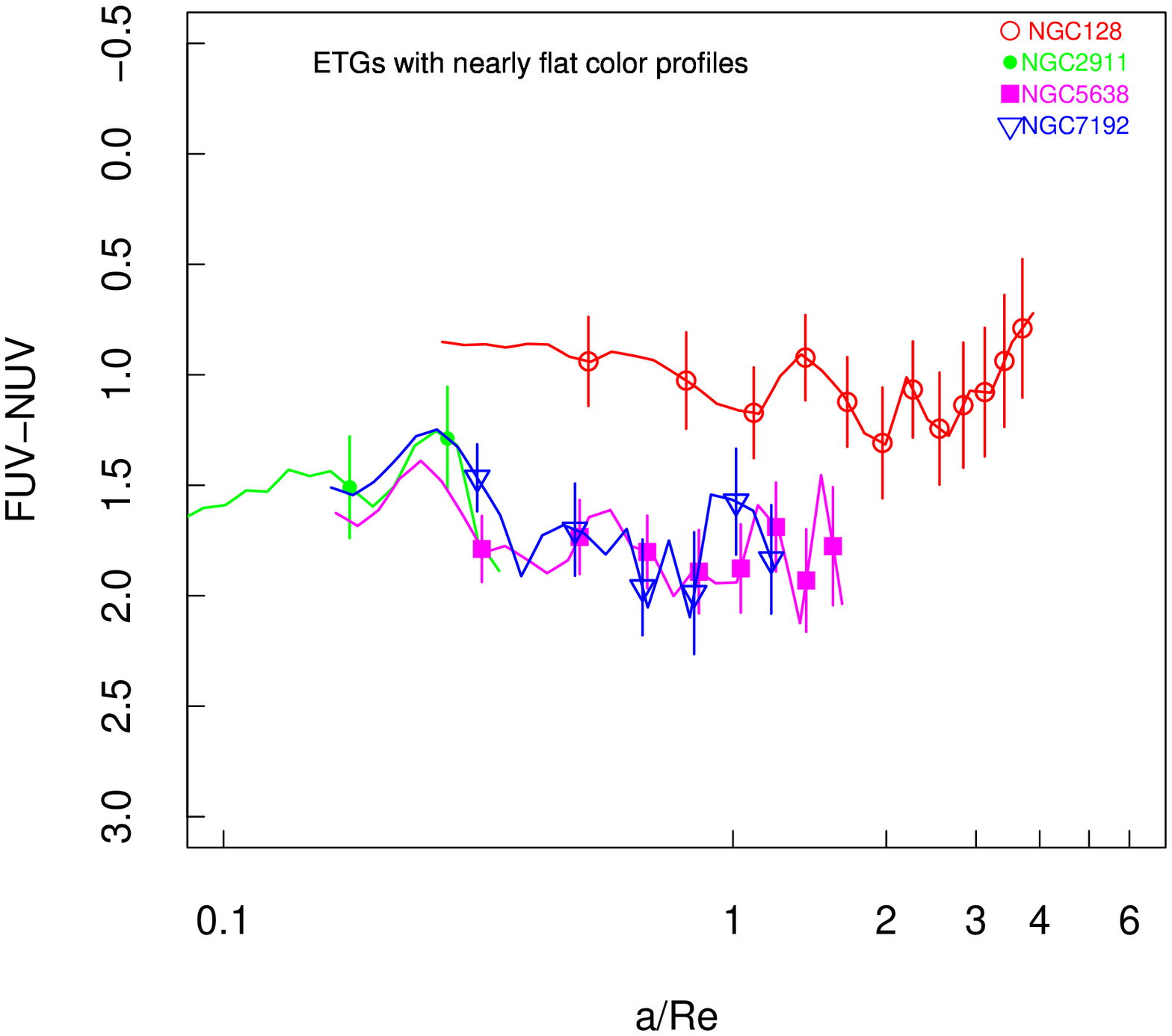,width=8.5cm}  
\hfill
\psfig{figure=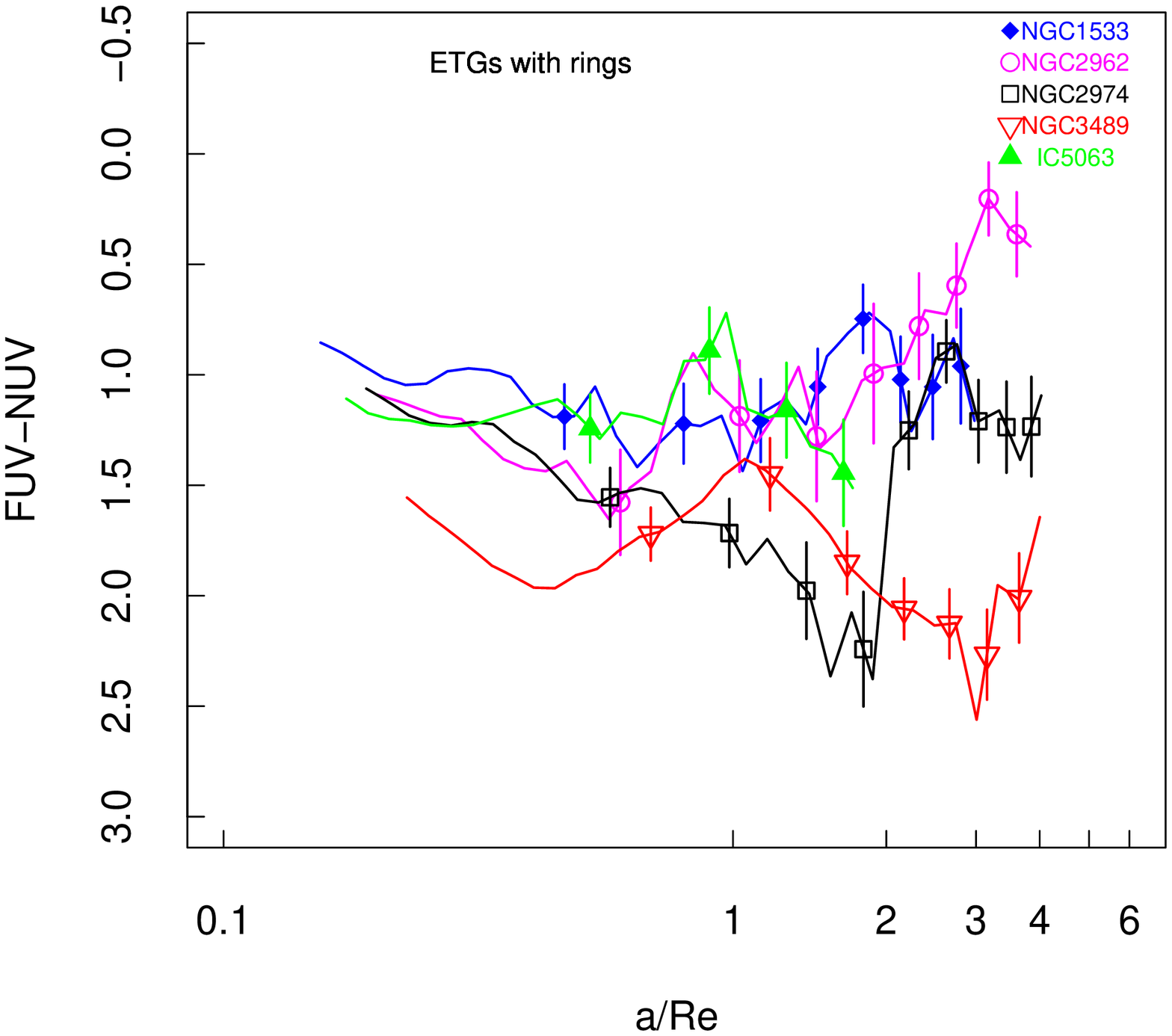,width=8.5cm}\\
\end{tabular}
\caption{UV radial colour profiles normalized to the optical effective radius. 
In the top panel we plot the UV radial profiles of the  rejuvenated  and 
{\it old} ETGs according to the luminosity-weighted ages estimated from 
the line-strength index analysis developed in Paper~III. The bottom left and
bottom right panels isolate a set of ETGs with nearly flat UV radial 
profiles and the set of ETGs showing a ring-like structure. } 
\label{fig5}
\end{figure*}

\begin{figure*}
\psfig{figure=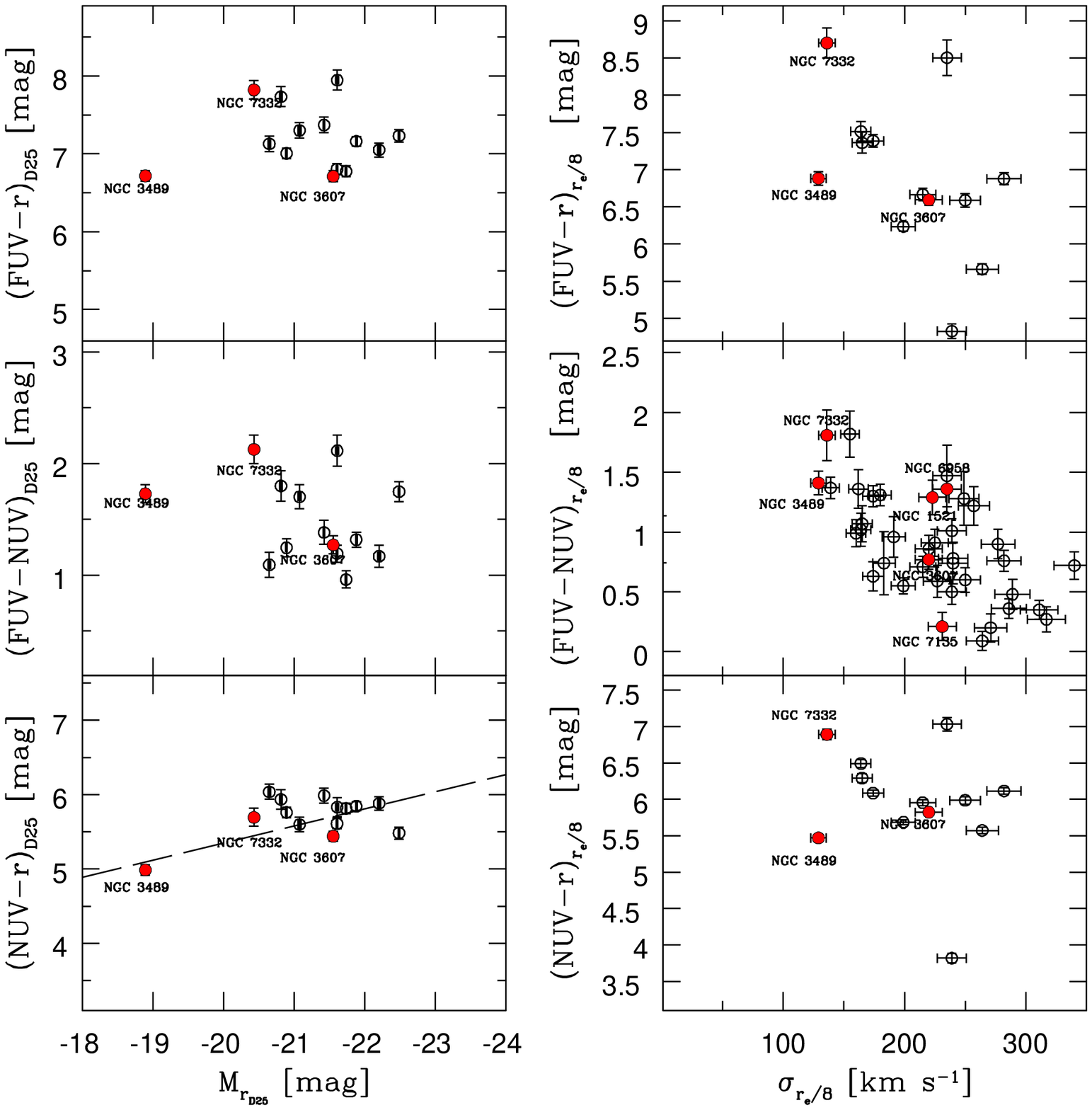,width=16.5cm} 
\caption{Left panels: UV - optical Color magnitude diagrams for a sub-sample of 14 
galaxies with SDSS r-band observations.  
 Magnitudes were computed within  D$_{25}$
(see Table 3).  The long-dashed line is the fit to 
the red sequence  at   0$< z <$ 0.05 derived by \citet{Yi05} . Right panels: colours  vs $\sigma$  at 
$r_8$/8  for the whole sample (except NGC 2962) and for the 14 galaxies with  SDSS data. 
Filled dots indicate ETGs whose luminosity-weighted 
age is $\la$4 Gyr from  Table~1.}  
 \label{fig6}
\end{figure*}

\subsection{The UV radial colour profiles}

Radial colour profiles  provide information about the presence 
of age/metallicity gradients in the stellar populations.
In Figure~\ref{fig5} we plot the (FUV-NUV) colour
as a function $a$/r$_e$, where a is the semi-major axis
and r$_e$ is the optical effective radius.
We stress that, while the UV colour profiles may extend out to several 
optical effective radii, the \ww ages, 
metallicities and $\alpha$-enhancements, 
derived in Paper~III, refer to apertures whose 
radius is only a fraction of r$_e$.

In the  upper panels of Figure~\ref{fig5} the (FUV-NUV) colour profiles 
of galaxies with a \ww age lower than 4 Gyr are shown in the left panel
and of galaxies older than 10 Gyr in the right panel, using age estimates from Paper~III
(see Table~1). Mean ages younger than 4 Gyr are likely due to recent star 
formation episodes superimposed on an old stellar population.
We exclude from these panels ETGs with rings which we
collect in the bottom right panel.

Figure~\ref{fig5} clearly shows that the (FUV-NUV) colour tends to 
become progressively redder from the centre outwards. This trend is consistent 
with the stellar populations becoming progressively more metal poor or younger 
(or a combination of both) with increasing distance from the galaxy centre.
In fact, the NUV light is dominated by turnoff stars.
On the other hand, the FUV emission of old `normal' stellar populations is
dominated by PAGB stars.
As a population gets younger or more metal poor, the turnoff becomes bluer
and more luminous, and emits more in the NUV; at the same time, the
contribution to the FUV from PAGB stars diminishes because,
in spite of the higher luminosity,
the duration of the PAGB phase gets much shorter (i.e., the fuel
decreases).
This causes  the FUV-NUV colour to become redder.
However, the situation could be more complex if  `anomalous'
sub-populations, such as those suggested to be responsible for the UV-upturn of
elliptical galaxies, were present. Several populations could develop
hot horizontal branch (HHB) and AGB Manqu\'e stars:
very helium-rich stars such as those recently found in globular cluster
multiple stellar populations \citep[e.g.][]{Piotto08};
ultra metal-rich stars \citep{Bressan94, Bertola95};
or normal metallicity stars with enhanced mass loss \citep{Yi97}.
Such populations could contribute, if not dominate, the FUV light.
The effects of such extreme populations  will be the subject of a forthcoming
investigation.

For populations younger than 1-2 Gyr (depending on the
metallicity) this trend is inverted, since the turnoff starts to contribute
significantly to the FUV band. Thus the (FUV-NUV) colour becomes steeply 
bluer with decreasing age.  
We conclude that the FUV-NUV colour profile cannot be used alone to
discriminate between negative age or metallicity gradients.
We know however from studies based on absorption line 
indices that strong negative metallicity gradients are present in ETGs.
Thus, the (FUV-NUV) colour trend may be, at least in part, due to a
metallicity effect. 

Three out of the five rejuvenated galaxies in the top left  panel 
of Figure~\ref{fig5} (NGC 1521, NGC 6958, NGC 7332) display redder colours
than the `old' galaxies. We notice that the {\it GALEX} (FUV-NUV) colour is not
very sensitive to reddening for Milky Way-type dust (an E(B-V)=0.3 
implies a $\Delta$m $\approx$0.2).
Thus, if the red colours were due to a reddening effect, we should have
an extinction as high as E(B-V)$\geq$0.5.
It is unlikely that such a strong extinction is present over several
galaxy effective radii. For this reason, we interpret the observed colours
as intrinsic. Lick indices do not show evidence for particularly 
low metallicities in the 5 galaxies.
On the contrary, the metallicities are quite high in all but NGC 7332  
consistent with their high velocity dispersions ($\sigma>$200 km~s$^{-1}$).
Thus, we suggest that the observed UV colours 
further support  the presence of young luminosity-weighted ages in
these galaxies.

In the bottom left panel of Figure~\ref{fig5} we plot separately
the profiles of four galaxies, namely NGC~128, NGC~2911, NGC~5638 and NGC~7192.  
Their luminosity-weighted ages range from 5.7$\pm$ 2.0  (NGC~7192 and
NGC~2911) to 9.7$\pm$1.7 Gyr (NGC~128) while their metallicities are
solar or twice solar (NGC~2911 and NGC~7192), the latter with 
a correspondingly  high central velocity dispersion.
Their (FUV-NUV) colour profiles tend to be nearly flat.
Three out of four of these galaxies show a very faint UV emission
extending out to 2 optical r$_e$. Only the  profile of NGC~128 extends 
up to 4 r$_e$ and it is very similar to that of NGC~7332 (top left panel), 
although much bluer ((FUV-NUV)$\approx$1) in the first galaxy 
vs. $\approx$2.2  in NGC~7332. Both galaxies are edge-on S0,
with nearly solar metallicity. 
 
In the right bottom panel of Figure~\ref{fig5}, the  colour profiles are
characterized by the presence of blue rings, which appear as
blue peaks in the (FUV-NUV) colours.
As discussed at the beginning of this section, an abrupt blueing
of the UV colours can be due to very recent star formation.
Notice however the presence of both rejuvenated (NGC~3489) and 
old (NGC~1533, NGC~2974) nuclei as indicated by the \ww
ages  derived in Paper~III in the r$_e$/8 region. 

We examined the behaviour of the central (FUV-NUV) colour vs. the `activity class'
(see Table~1, column 11). Although most of our galaxies have a LINER nuclei,  
Seyfert,  Composite and Inactive nucleus are present (see Paper~IV). 
We find that the central UV colour is totally uncorrelated with the activity class. 
The bluest  (FUV-NUV) colours are  seen in NGC~3258 and NGC~4552 nuclei,
classified in  Paper~IV as Composite,  from diagnostic diagrams
\citep{Kewley01,kewley06}. NGC~3258 (4.5$\pm$0.08 Gyr) and NGC~4552 
(6.0$\pm$1.4) could have experienced a recent star formation event.   
 A similar result is found by \cite{Suh10} on a large   ETG
sample drown from the SDSS DR6. 
Their analysis of the radial color gradient of  ETGs,
 shows that the so-called blue-cored galaxies, possible 
rejuvenated ETGs, occupy both the star-formation and Seyfert regions
in the classical [OIII]/H$\beta$ vs. [NII]/H$\alpha$ BTP diagram \citep{Baldwin81}. 
These authors suggest that most blue-cored ETGs have 
centrally concentrated star formation and a smaller fraction 
could be associated with Seyfert activity.

\subsection{Star formation and the large scale HI distribution}

The  FUV flux and the HI distributions correlate in star forming regions 
\citep[see e.g.][]{Neff05, Thilker05}. In this context, we investigated in the literature  
the presence and the morphology of the large scale HI distribution in the present ETG sample. 

Only a few objects have been observed. The HI content of NGC~1052 has been observed 
with the VLA by \citet{vanGorkom86}. Their Figure 5 shows an irregular HI structure,
elongated roughly perpendicularly to the galaxy major axis. 
In the southwest region of NGC~1052,  the HI distribution shows what 
appears to be a tidal tail, suggesting that the gas may have been acquired  
about 10$^9$ years ago.
We notice that in NGC~1052 there is no correlation between the HI distribution  and the 
far UV emission.  Similar cases of tail-like HI distributions, decoupled from the 
Far UV emission, have been detected in some  shell galaxies, namely
NGC~2865, NGC~5018 and NGC~7135 \citep{Rampazzo07}.

\citet{Werk10} show the large scale HI distribution  in the field of 
NGC~1533, one of our ETGs showing a blue outer ring. The HI distribution is
connected to a companion galaxy, IC~2038. Four  outer-HII regions are 
detected, powered by young, low mass OB associations according to
\citet{Werk10} (see their Figure 7). 
Their FUV-NUV colour, in the range $-0.79\pm 0.37\leq (FUV-NUV) \leq 0.03\pm 0.28$,
is bluer than the knotty regions in the galaxy ring, which we will 
discuss and model in a forthcoming paper \citep{Marino10}. 

The HI distribution has been investigated also in NGC~4374 
\citep{Li01} and in NGC~5846 \citep{Zabludoff01}. HI 
is detected in neither, although both have HI rich companions. 

\subsection{UV-optical colour vs. absolute magnitude and velocity dispersion relations}
  
The left panels of Fig. \ref{fig6} present UV-optical colour-magnitude
relations (CMRs) of the 14 galaxies for which SDSS data are available
(see Table 3 and Appendix B).
Magnitudes were computed within D$_{25}$.
No clear correlations are observed
in the (FUV$-$NUV) vs M$_r$ and
(FUV$-$r) vs M$_r$ planes, while in the (NUV$-$r) vs M$_r$ plane the positive
trend is mainly driven by one object (NGC 3489) at relatively faint
magnitudes.
In this plane, our galaxies are consistent with the local universe
(0$<$z$<$0.05) red sequence derived by \citet{Yi05} ((-0.23$\pm$0.30)$\times$ M$_r$ + 0.75).
We also notice that rejuvenated galaxies are mixed with older ETGs in
these CMRs.  

In the right panels of Fig.6,
we plot the UV-optical colours derived within
$r_e/8$ versus the central velocity dispersion $\sigma_{r_e/8}$.
A significant correlation is observed only in the FUV$-$NUV vs $\sigma$
plane, but in this case the sample is larger (39 galaxies).
The Spearman correlation coefficient is $r_s=-0.62$ with N$=$37 degrees of
freedom, indicating that an anti-correlation exists with a probability $>$
0.99.
This result is in agreement with \citet{Donas07, Jeong09}.

Studies based on narrow-band indices have shown that the stellar
populations in
ETGs tend to become progressively more metal rich and older
with increasing velocity dispersion
(i.e., increasing galaxy mass) \citep[e.g][]{Thomas05, Clemens06, Clemens09}.
 Both effects could be responsible for the observed FUV$-$NUV trend since
in general this colour becomes bluer for more metal rich and older passive
populations (see discussion in Section 4.3). 
 
\section{Summary and Conclusions}

We have obtained  {\it GALEX} NUV and FUV aperture and
surface photometry of 40 ETGs whose  
emission line properties have been presented in  Paper~III 
and Paper~IV. In Appendix~B we also provide the SDSS $r$-band surface 
photometry of 14 of the above ETGs. 

We used UV and optical imaging  to study 
mechanisms driving both  secular and/or external ETG 
evolution such as bars and recent accretion episodes, respectively, 
as suggested by the young \ww ages resulting from
the line strength indices analysis (see Paper~III). 

We find the following results:
\newcounter{count1}
\begin{list}{\arabic{count1}. }{\usecounter{count1}}

\item In general, the (FUV$-$NUV) radial profiles become redder with
galactocentric distance  \citep[see also][]{Jeong09}.
This property is common to both rejuvenated ($\la$ 4 Gyr) and old ETGs.
The trend could be due to age  and/or metallicity gradients.
The presence of strong metallicity gradients in these ETGs as revealed by
the analysis of narrow-band indices in Paper~III suggests that the
radial (FUV$-$NUV) profiles are due, at least in part, to radial
metallicity variations. The role of the metallicity could be  
dominant if, as found by Clemens et al 2009,
the age increases with galactocentric distance.
The study of moderate redshift (0.4 $< z <$ 1.5) ETGs performed 
by GOODS (Great Observatory Origins Deep Survey) \citep{Ferreras09}
suggest that redshift evolution of the observed color gradients 
is incompatible with a significant variation in stellar age within
each galaxy and that the local observations of radial color gradients
mostly correspond to a range of metallicities.

\item Relevant  exceptions to what seems the `normal' (FUV-NUV)
colour profile are ring/arm-like galaxies.
The NUV and FUV images of NGC~1533, NGC~2962, NGC~2974, NGC~3489
and IC~5063 show that ring and/or arm-like structures are bluer
than the galaxy body. 
Kinematical models suggest that rings have an internally driven origin and
are closely associated with the evolution of the bar. The gas may be driven 
both in the centre and  in the ring region where accumulates and may activate 
star formation. The young luminosity-weighted 
age of NGC~3489 and its AGN-like activity may be explained in the framework 
of a gas inflow toward the galaxy centre. In Paper~IV we found evidence that 
star formation and AGN activity are closely connected in time.
We interpret the abrupt blueing of the  UV colour in the ring/armlike structure as due to
recent star formation. Detailed models are planned in a subsequent work. 
 We consider these ETGs as possible 
candidates of ongoing secular evolution. 

\item  Shell galaxies are a prototypical example of the externally driven evolution
of ETGs. Out of the seven ETGs in the sample with stellar shells detected in the optical, 
NGC~7135 is the only shell galaxy showing a shell structure in the NUV and FUV images. 
\citet{Rampazzo07} suggest that this galaxy could have had 
a recent nuclear star formation episode as implied by the low \ww
  age, likely triggered by the accretion event that formed the shell.

\item The Sersic profile shape parameter $n$ shows much variation in UV (1 $< n < 16$) as in the
optical and near infrared bands.  S0s tend to have values of $n<5$
in both NUV and FUV, as already noticed in the optical.
The Sersic index $n=4$ is a sort of  watershed as noted in the
optical by \citet{Kormendy09}. Most ETGs with $n>4$ 
have [$\alpha$/Fe] $\geq$ 0.15, implying a short star-formation 
time scale. ETGs with n=3$\pm$1 have a large spread in [$\alpha$/Fe], 
including very low values indicative of a more prolonged star formation. 
Most of these galaxies are S0s, i.e truly
disk galaxies, which, as noticed in Paper~III, generally appear more
rejuvenated with respect to Es. These galaxies have also, in general, 
a low central velocity dispersion.   We suggest that the Sersic index
vs.  [$\alpha$/Fe] trend reflects the velocity dispersion (i.e. mass)
$\alpha$-enhancement relation. 

\item A significant correlation between the (FUV$-$NUV) colour and $\sigma$
is observed, in agreement with previous studies \citep{Donas07,
Jeong09}. This trend is likely to be driven by a combined effect
of the age-$\sigma$ and metallicity-$\sigma$ relations found
from narrow band indices. A more detailed investigation on the origin
of this relation will be performed in a forthcoming paper.

\end{list}   

A detailed analysis of the UV colour vs. line-strength indices and of
the ETGs Spectral Energy distribution from UV to optical and MIR 
will be presented in forthcoming papers.

\section*{Acknowledgments}
We acknowledge Mauro D'Onofrio for his support, valuable
discussions and comments.
AM acknowledges support from Italian Scientist 
and Scholars in  North America Foundation (ISSNAF) through 
an ISSNAF fellowship in Space Physics and Engineering, sponsored 
by Thales Alenia Space. RR, AB, FA, LMB acknowledge the ASI-INAF support
through   contract  I/016/07/0. 
{\it GALEX} is a NASA Small Explorer, launched in April
2003. {\it GALEX} is operated for NASA by California Institute of
Technology under NASA contract  NAS-98034. 
This work was partly supported by NASA grant NNX07APO8G.
This research has made
use of  SAOImage DS9, developed by Smithsonian Astrophysical 
Observatory and of the NASA/IPAC Extragalactic Database (NED) which 
is operated by the Jet Propulsion Laboratory, California Institute of
Technology, under contract with the National Aeronautics and Space
Administration. {\tt IRAF} is distributed
by the National Optical Astronomy Observatories, which are operated
by the Association of Universities for Research in Astronomy, Inc.,
under cooperative agreement with the National Science Foundation.
We acknowledge the usage of the HyperLeda database (http://leda.univ-lyon1.fr).\\
{\it Facilities:} {\it GALEX}, SDSS

\appendix

\section{UV and SDSS description of individual galaxies}  

We briefly comment here on  the {\it GALEX} and SDSS
surface brightness profiles of the ETGs presented in this paper.
We also list morphological and kinematic peculiarities which
could be relevant for understanding the galaxy evolution. 
For a detailed information about the \ww age,
metallicity and $\alpha$-enhancements as well as about
emission line properties we cross-refer to Papers~III and Papers~IV
(these data are summarized in Table~1).
The basic  information about ETGs is provided in Papers~I, II (on line material) 
and IV (Appendix A).  

{\it NGC~128}~~~~~ This edge-on S0 is a LINER. The (FUV-NUV)  colour profile
is flat within the errors (average $\approx$1.02$\pm$0.14).  The average
 Sersic index is $\langle n_{ave} \rangle$ =2.74$\pm$0.66.
Gas and stars counter rotate in this about 10 Gyr old fast rotator.

{\it NGC~777} ~~~~~  The (FUV-NUV) colour profile drops from $\approx$0.5 in 
the centre of the galaxy to $\approx$1.5 in the outskirts. This slow rotator, 
Seyfert-like galaxy has a young (5.4$\pm$2.1) \ww 
age in the nucleus.   The average Sersic index is $\langle n_{ave} \rangle$= 6.14$\pm$2.16:
the quality of the fit is indeed quite poor.

{\it NGC~1052} ~~~~~  This old ($\approx$14 Gyr), fast rotating galaxy is 
considered a prototypical LINER.  A gas vs. gas counter rotation is 
reported for this galaxy. As in previous galaxy the (FUV-NUV) colour profile 
becomes red  going from the centre ($\approx$0.75) 
to the outskirts ($\approx$1.75). (NUV-$r$) becomes slightly 
bluer  from $\approx$0.55 to 0.5 while (FUV-$r$) redder from the 
centre $\approx$6.2 to 6.9. The average Sersic index is 
$\langle n_{ave} \rangle$= 4.53$\pm$2.83. The large spread in the $n$ indices
does not depend from the quality of the fit.

Appendix A is available in its entirety in the online edition of the Journal. 
A small portion is shown here for guidance.
 
\section{SDSS surface photometry of a galaxy sub-sample}  

Figure~A1 displays composite SDSS images available 
for  14 ETGs in our sample.
Figure~A2
shows the surface brightness profiles 
in the $r$ band and the 
(FUV-$r$) and (NUV-$r$) colour profiles.


 \begin{figure*}
\psfig{figure=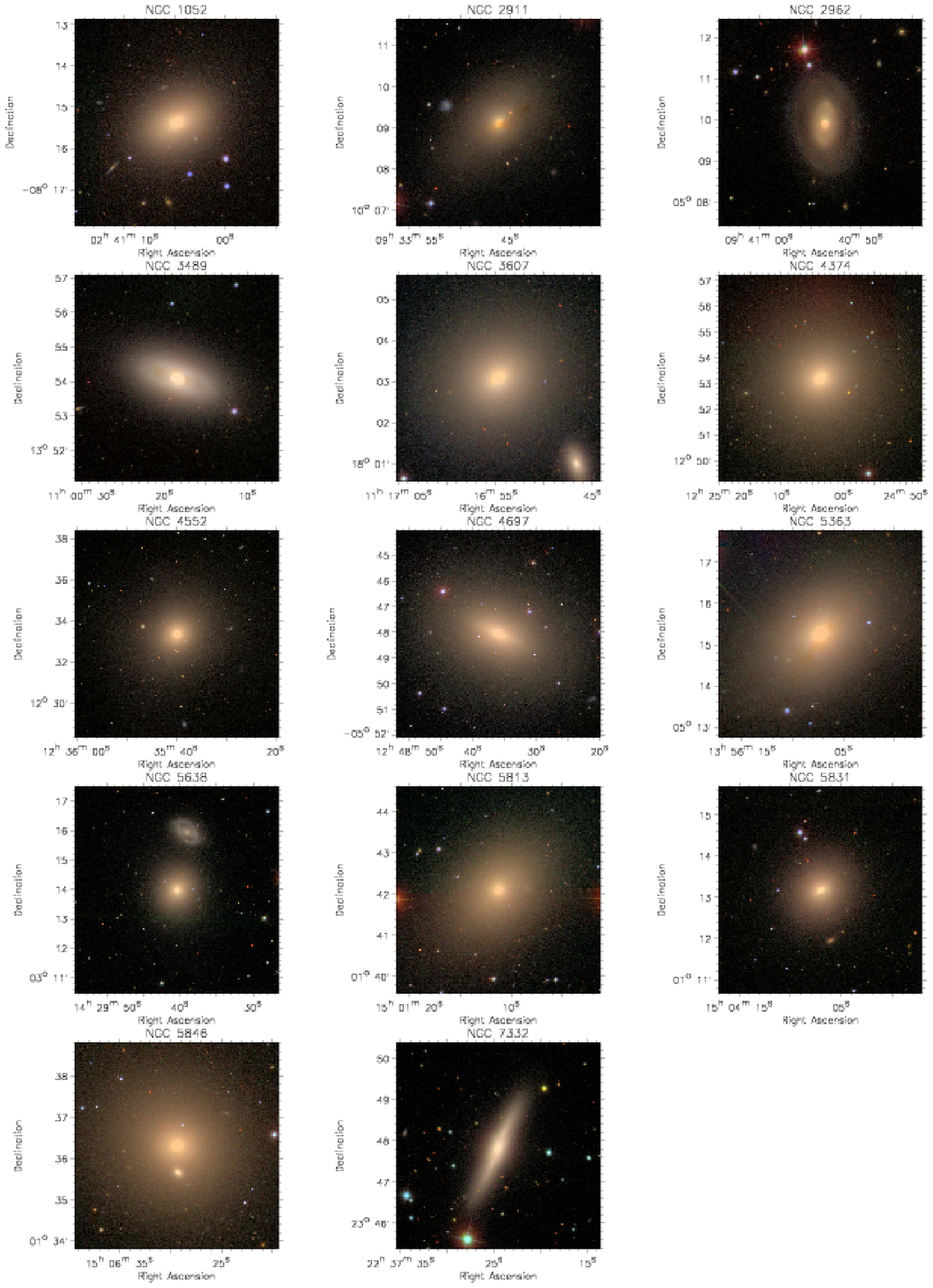,width=15cm} 
\caption{False colour composite SDSS images ({\it g} blue, {\it r} green,  {\it i} red) available for 14 ETGs of the sample.} 
  \label{SDSS}
\end{figure*}

\begin{figure*}
\psfig{figure=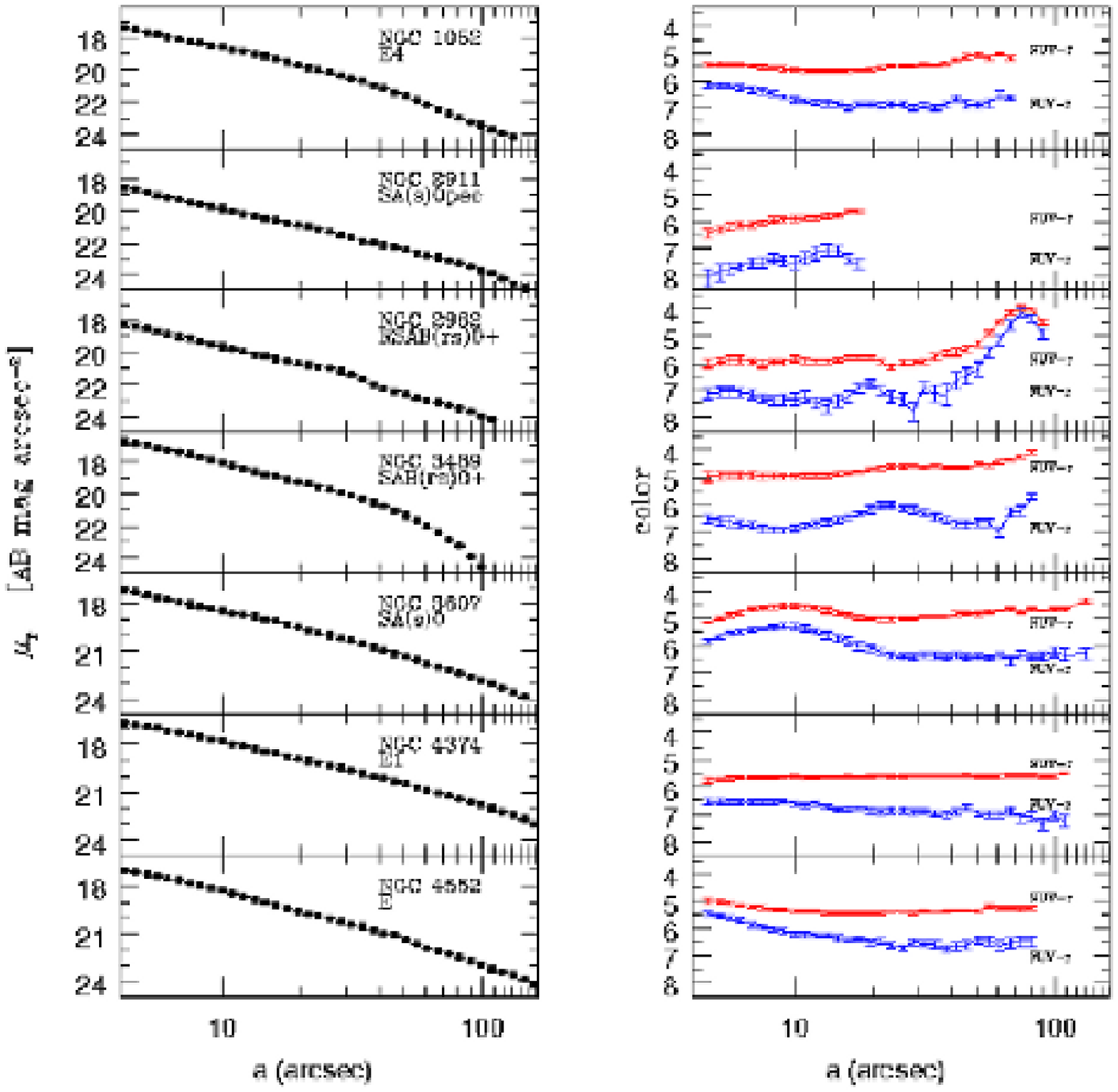,width=19.2cm} 
 \caption {From top to bottom: (left panels) Luminosity profiles in the   SDSS $r$ band,
(right panels) (FUV-r) and (NUV-r) colour profiles along the semi-major axis of the fitted ellipse.} 
 \end{figure*}

\begin{figure*}
\psfig{figure=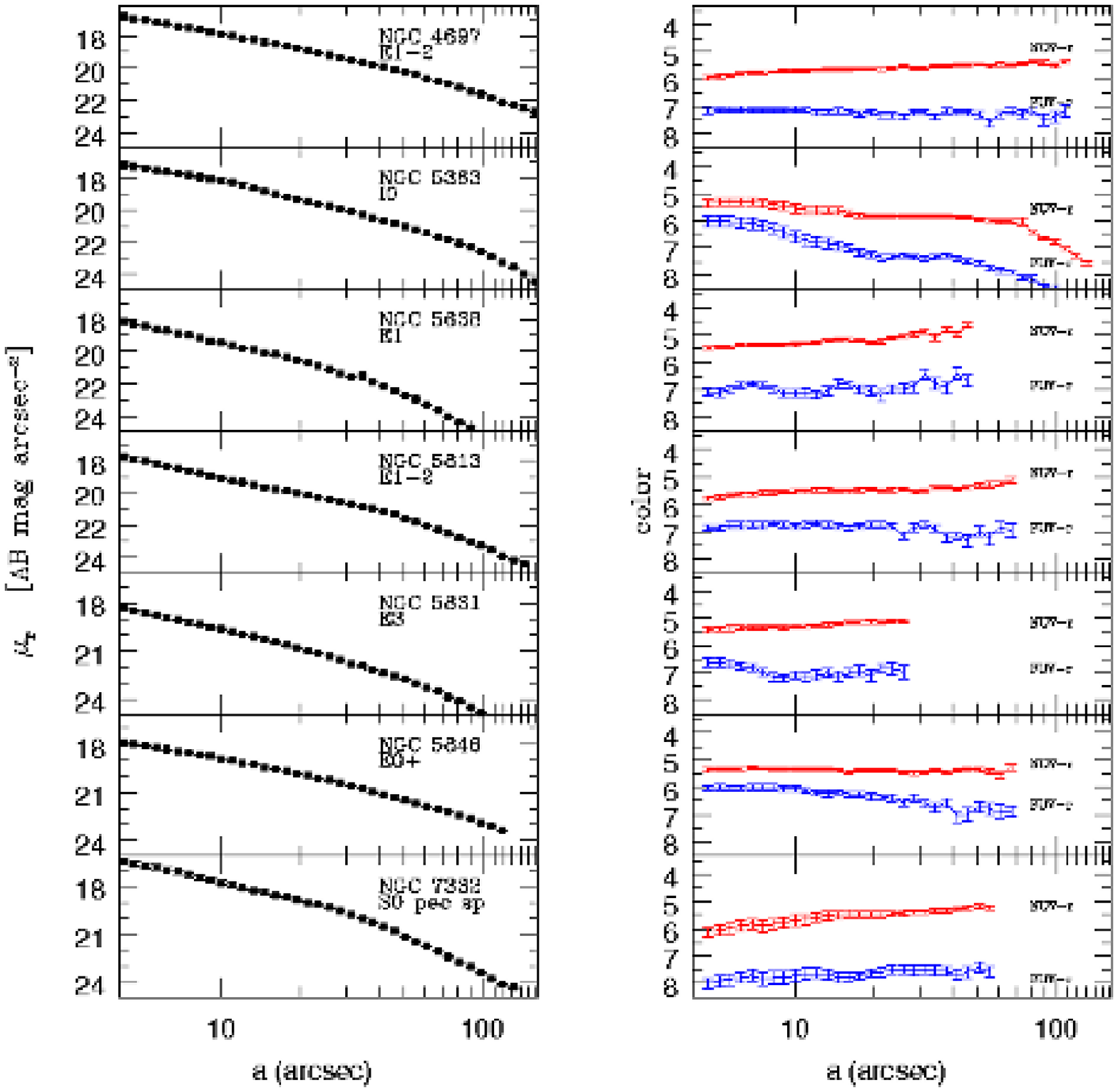,width=19.2cm} 
\addtocounter{figure}{-1}
\caption{Continued.} 
 \end{figure*}

\label{lastpage}

\end{document}